\documentclass[journal, 10pt]{IEEEtran}
\usepackage[T1]{fontenc}
\usepackage[latin9]{inputenc}
\usepackage{array}
\usepackage{float}
\usepackage{amsmath}
\usepackage{amsthm}
\usepackage{amssymb}
\usepackage{graphicx}
\usepackage[unicode=true,
 bookmarks=true,bookmarksnumbered=true,bookmarksopen=true,bookmarksopenlevel=1,
 breaklinks=false,pdfborder={0 0 0},pdfborderstyle={},backref=false,colorlinks=false]
 {hyperref}
\hypersetup{pdftitle={Your Title},
 pdfauthor={Your Name},
 pdfpagelayout=OneColumn, pdfnewwindow=true, pdfstartview=XYZ, plainpages=false}

\makeatletter

\floatstyle{ruled}
\newfloat{algorithm}{tbp}{loa}
\providecommand{\algorithmname}{Algorithm}
\floatname{algorithm}{\protect\algorithmname}

\theoremstyle{plain}
\newtheorem{thm}{\protect\theoremname}
\theoremstyle{plain}
\newtheorem{prop}{\protect\propositionname}

\usepackage[caption=false,font=footnotesize]{subfig}
\usepackage{algpseudocode}

\algdef{SE}{Begin}{End}{\textbf{begin}}{\textbf{end}}
\algnewcommand\algorithmicforeach{\textbf{for each}}
\algdef{S}[FOR]{ForEach}[1]{\algorithmicforeach\ #1\ \algorithmicdo}
\newcommand{\Break}{\State \textbf{break}}
\makeatother

\providecommand{\propositionname}{Proposition}
\providecommand{\theoremname}{Theorem}

\begin{document}
\title{Toward an Automated Auction Framework for Wireless Federated Learning
Services Market}
\author{Yutao Jiao, Ping Wang, \IEEEmembership{Senior Member,~IEEE,} Dusit
Niyato, \IEEEmembership{Fellow,~IEEE}, Bin Lin,\\and Dong In Kim,
\IEEEmembership{Fellow,~IEEE}\thanks{Yutao Jiao and Dusit Niyato are with the School of Computer Science
and Engineering, Nanyang Technological University, Singapore. Ping
Wang is with the Lassonde School of Engineering, York University,
Canada. Bin Lin is with the College of Information Science and Technology,
Dalian Maritime University, China. D. I. Kim is with the Department
of Electrical and Computer Engineering, Sungkyunkwan University, Suwon
16419, South Korea.}}
\maketitle

\begin{abstract}
In traditional machine learning, the central server first collects the data
owners' private data together and then trains the model. However, people's
concerns about data privacy protection are dramatically increasing.
The emerging paradigm of federated learning efficiently builds machine
learning models while allowing the private data to be kept at local
devices. The success of federated learning requires sufficient data
owners to jointly utilize their data, computing and communication
resources for model training. In this paper, we propose an auction
based market model for incentivizing data owners to participate in
federated learning. We design two auction mechanisms for the federated
learning platform to maximize the social welfare of the federated
learning services market. Specifically, we first design an approximate
strategy-proof mechanism which guarantees the truthfulness, individual
rationality, and computational efficiency. To improve the social welfare,
we develop an automated strategy-proof mechanism based on deep reinforcement
learning and graph neural networks. The communication traffic congestion
and the unique characteristics of federated learning are particularly
considered in the proposed model. Extensive experimental results demonstrate
that our proposed auction mechanisms can efficiently maximize the
social welfare and provide effective insights and strategies for the
platform to organize the federated training.
\end{abstract}

\maketitle

\section{Introduction \label{sec:Introduction} }

\IEEEPARstart{C}{}urrently, there are nearly $7$ billion connected
Internet-of-Things (IoT) devices\footnote{https://iot-analytics.com/state-of-the-iot-update-q1-q2-2018-number-of-iot-devices-now-7b/}
and $3$ billion smartphones around the world. The devices continuously
generate a large amount of fresh data. The traditional data analytics
and machine learning requires all the data to be collected to a centralized
data center/server, and then used for analysis or produce effective
machine learning models. This is the actual practice now conducted
by giant AI companies, including Amazon, Facebook, Google, etc.. However,
this approach may raise concerns regarding the data security and privacy.
Although various privacy preservation methods have been proposed,
such as differential privacy~\cite{Abadi2016} and secure multi-party
computation (MPC)~\cite{Goldreich1998}, a large proportion of people
are still not willing to expose their private data which can be inspected
by the server. This discourages the development of advanced AI technologies
as well as new industrial applications. Motivated by the increasing
privacy concern among data owners, Google introduced the concept of
the \emph{federated learning (FL)}~\cite{BrendanMcMahan2016}. The
FL is a collaborative learning scheme that distributes the training
process to individual users which then collaboratively train the shared
model while keeping the data on their devices, thus alleviating the
privacy issues. 

A typical FL system is composed of two entities, including the \emph{FL
platform} and the \emph{data owners}. Each data owner, e.g., mobile
phone user, has a set of private data stored at its local device.
The \emph{local data} are used to train a local machine learning model
where the initial model and hyper-parameters are preset by the FL
platform. Once the local training is completed, each data owner just
sends the trained model to the FL platform. Then, all received \emph{local
models} are aggregated by the FL platform to build a \emph{global
model}. The training process iterates until achieving the target performance
or reaching the predefined number of iterations. Federated learning
has three distinctive characteristics~\cite{Park2019, Lim2019}: 
\begin{enumerate}
	\item A massive number of distributed FL participants are independent and uncontrollable, which is different from the traditional distributed training at a centralized data center.
	\item The communication among devices, especially through the wireless
	channel, can be asymmetric, slow and unstable. The assumption of a
	perfect communication environment with a high information transmission
	rate and negligible packet loss is not realistic. For example, the
	Internet upload speed is typically much slower than download speed.
	Some participants may consequently drop out due to disconnection to
	the Internet, especially using the mobile phone through congested
	wireless communication channels~\cite{Kairouz2019}.
	\item The local data is not independent and identically distributed (Non-IID), which significantly affects the learning performance~\cite{Zhao2018, Liu2019}. Since data owners' local data cannot be accessed and fused by the FL platform and may follow different distributions, assuming all local datasets are IID is impractical.
\end{enumerate}

As implied by the first characteristic above, an important prerequisite
for a successful FL task is the participation of a large base of data
owners that contribute sufficient training data. Therefore, establishing
an FL services market is necessary for the sustainable development
of the FL community. We propose an auction based market model to facilitate
commercializing federated learning services among different entities.
Specifically, the FL platform first initiates and announces an FL
task. When receiving the information of the FL tasks, each data owner
determines the service value by evaluating its local data quality
and the computing and communication capabilities. Then, data owners
report their types including bids representing the services value
and their resources information to the FL platform. According to the
received types, the platform selects a set of FL workers from data
owners and decides the service payments. Finally, the FL platform
coordinates the selected FL workers to conduct model training.

In this paper, we mainly investigate the federated learning in the
wireless communication scenario and design applicable auction mechanisms
to realize the trading between the FL platform and the data owners.
From the system perspective, we aim to maximize their total utility,
i.e., social welfare. For an efficient and stable business ecosystem
of the FL services market, there are several critical issues about
FL task allocation and pricing. First, which data owner can participate
in the federated training as an FL worker? Due to the unique features
listed above, the FL platform should consider data owners' reported
data size and non-IID degree of data. Also, the limited wireless spectrum
resource need to be reasonably allocated since the large population
of participated data owners may exacerbate the communication congestion.
Second, how to set reasonable payments for data owners such that they
can be incentivized to undertake the FL tasks? Auction is an efficient
method for pricing and task allocation~\cite{Krishna2009}. The payment
amount should satisfy individual rationality, which means there is
no loss to data owners from trading. We should also consider how to
make data owners truthfully expose their private types. The truthfulness
property can stabilize the market, prevent possible manipulation and
may significantly reduce the communication overhead and improve the
learning efficiency. The major contributions of this paper can be
summarized as follows:
\begin{itemize}
\item Based on real-world datasets and experiments, we define and verify
a data quality function that reflects the impacts of local data volume
and distribution on the federated training performance. The earth
mover\textquoteright s distance (EMD)~\cite{Zhao2018} is used as
the metric to measure the non-IID degree of the data. Moreover, we
consider the wireless channel sharing conflicts among data owners.
\item We propose an auction framework for the wireless federated learning
services market. From the perspective of the FL platform, we formulate
the social welfare maximization problem which is a combinatorial NP-hard
problem. 
\item We first design a reverse multi-dimensional auction (RMA) mechanism
as an approximate algorithm to maximize the social welfare. To further
improve the social welfare and the efficiency, we novelly develop
an automated deep reinforcement learning based auction (DRLA) mechanism
which is integrated with the graph neural network (GNN). According
to the data owners' requested wireless channels, we construct a conflict
graph for the usage of GNN. Both mechanisms, i.e., RMA and DRLA, are
theoretically proved to be strategyproof, i.e. truthful and individually
rational. 
\item Demonstrated by our simulation results, the proposed auction mechanisms
can help the FL platform make practical trading strategies to efficiently
coordinate data owners to invest their data and computing resources
in the federated learning while optimizing the social welfare of the
FL services market. Particularly, the automated DRLA mechanism shows
significant improvement in social welfare compared with the RMA mechanism.
\end{itemize}
To the best of our knowledge, this is the first work that studies
the auction based wireless FL services market and applies the GNN
and deep reinforcement learning (DRL) in the design of a truthful
auction mechanism to solve a combinatorial NP-hard problem.

The rest of this paper is organized as follows. Section~\ref{sec:Related_Work}
reviews related work. The system model of the FL services market and
the social welfare maximization problem are introduced in Section~\ref{sec:FL_Market}.
Section~\ref{sec:Reverse_multidimenstional_auction} proposes the
designed reverse multi-dimensional auction mechanism. In Section~\ref{sec:DL_auction},
the automated auction mechanism based on GNN and DRL is presented
in detail. Section~\ref{sec:Experimental-and-simulation} presents
and analyzes simulation results based on real-world and synthetic
datasets. Finally, Section~\ref{sec:Conclusion} concludes this paper.

\begin{figure*}[tbh]
\begin{centering}
\includegraphics[width=0.62\paperwidth]{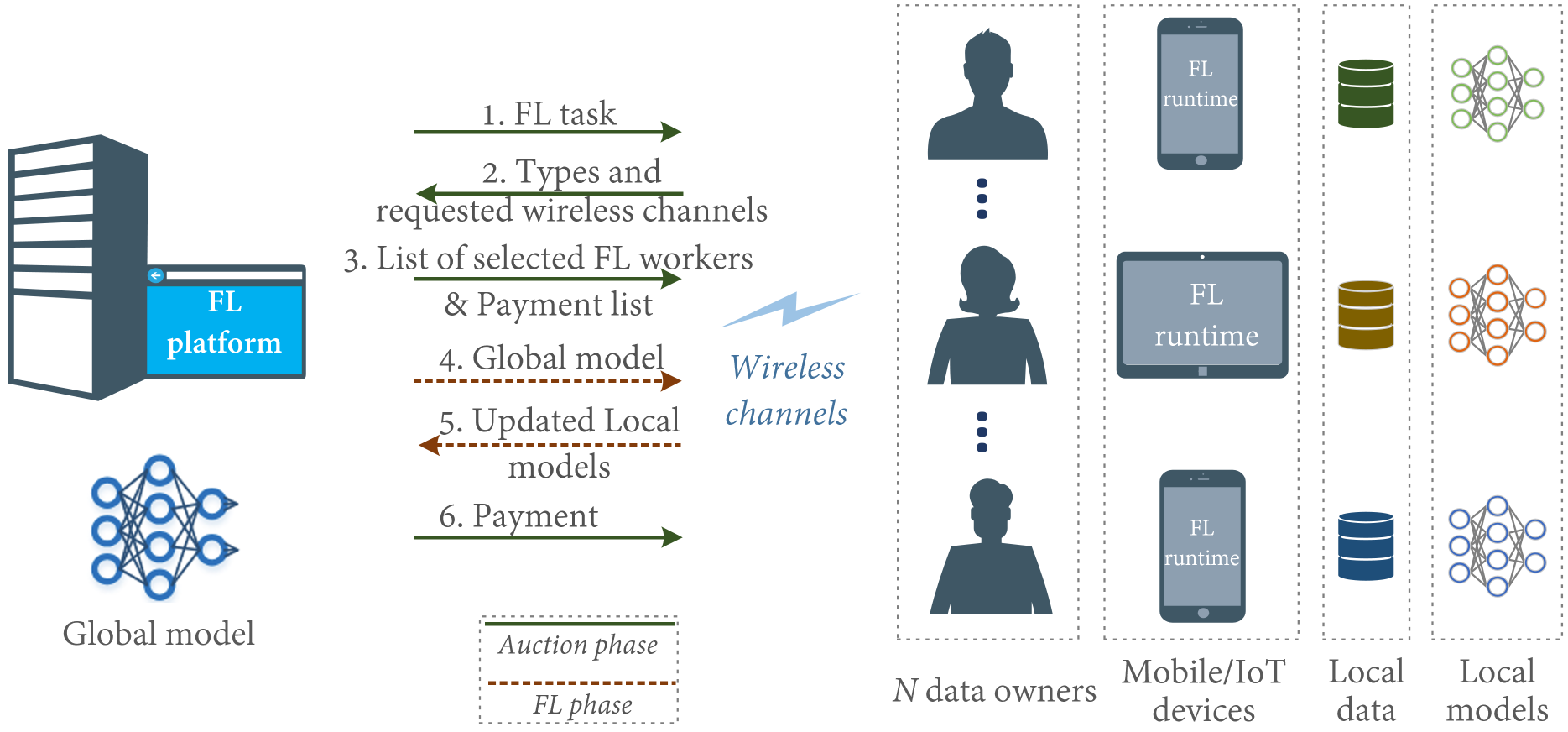}
\par\end{centering}
\caption{Federated learning services market.\label{fig:FL_market}}
\end{figure*}

\section{Related Work\label{sec:Related_Work} }

Due to the resource constraints and the heterogeneity of devices,
some papers have discussed the optimal allocation of the resources
and tasks to improve the efficiency of federated training. The relevant
issues mainly include client selection, computation offloading and
incentive mechanism. The authors in~\cite{Nishio2019} designed a
protocol called FedCS. The FedCS protocol has a resource request phase
to gather information such as computing power and wireless channel
states from a subset of randomly selected clients, i.e., FL workers.
To tradeoff the accuracy and efficiency, the FL platform optimally
selects a set of clients that are able to punctually finish the local
training. Compared with the protocol that ignores the client selection,
the FedCS can achieve higher performance. Besides improving the training
efficiency, the authors in~\cite{Mohri2019,Li2019a} discussed the
fairness issue that if a protocol selects the clients by the computing
power, the final trained model would more cater to the data distribution
of clients with high computational capability. Based on the original
federated averaging (FedAvg) algorithm~\cite{Konevcny2015}, a $q$-FedAvg
training algorithm was proposed in~\cite{Li2019a} to give the client
with low performance a higher weight in optimizing the objective function.
For computation offloading, the authors in~\cite{Wang2018a} combined
the DRL and the FL to optimally allocate
the mobile edge computing (MEC) resources. The client can use the
DRL to intelligently decide whether to perform the training locally
or offload it to the edge server. The simulation results showed that
the DRL based approaches can achieve similar average utilities in
FL and centralized learning. With respect to the incentive mechanism
design, the authors in~\cite{Feng2018} proposed a Stackelberg game
model to investigate the interactions between the server and the mobile
devices in a cooperative relay communication network. The mobile devices
determine the price per unit of data for individual profit maximization,
while the server chooses the size of training data to optimize its
own profit. The simulation results demonstrate that the interaction
can finally reach an equilibrium, and the cooperative communication
scheme can reduce the congestion and improve the energy efficiency.
In a similar setting of \cite{Feng2018}, the authors in~\cite{Kang2019}
proposed a contract theory method to incentivize the mobile devices
to take part in the FL and contribute high-quality data. The mobile
users can only choose the contract matching their own types to maximize
the utility. However, the above incentive mechanisms did not consider
the non-IID data or the wireless channel constraints which are taken
into account in this paper. 

Different from the Stackelberg game and contract theory, the auction
mechanism allows the data owner to actively report its type and has been applied in various application scenarios~\cite{GaoTWC2019}. Thus,
the FL platform can sufficiently understand their status and requests
to optimize the target performance metric, such as the social welfare
of the market or the platform's revenue. To design a new auction mechanism
for higher performance or other properties that manually designed
auction mechanism cannot realize, the automated mechanism design~\cite{Conitzer2004,Sandholm2015} assisted by machine learning techniques is gaining popularity. In~\cite{Duetting2019}, the authors used the multi-layer neural network to model an auction
with the guarantee of individual rationality (IR) and incentive compatibility\footnote{In this paper, truthfulness and incentive compatibility are used interchangeably.}
(IC). 
The proposed deep learning based framework successfully recovered
all known analytical solutions to classical multi-item auction settings,
and discovered new mechanisms for settings where the optimal analytical
solution is unknown. In~\cite{JiaoTVT2019}, the authors proposed a strategyproof mechanism to deploy the mobile base station based on the deep learning technique, which significantly improved the  social welfare of the wireless powered spatial crowdsourcing system. The study of using DRL to solve combinatorial problems over the graph was initialized in~\cite{Khalil2017}. The
authors first calculated the graph embedding and then trained a deep
Q network to optimize several classical NP-hard problems in a greedy
style. Since the wireless channel conflicts among the data owners
are represented by a conflict graph in this paper, we propose an automated
auction mechanism based on DRL and GNN to optimize the social welfare
of FL services market while meeting the requirement of IC and IR.

\section{System Model: Federated Learning Services Market\label{sec:FL_Market}}

\subsection{Preliminary Knowledge of Federated Learning\label{subsec:Preliminaries_of_FL}}

As illustrated in Fig.~\ref{fig:FL_market}, we focus on a representative
monopoly FL services market structure which consists of one \emph{FL
platform} and a community of \emph{$N$ data owners} $\mathcal{N}=\left\{ 1,\ldots,N\right\} $.
The \emph{platform} performs publishing the FL task and selecting
data owners as \emph{FL workers}. Each data owner $i$ maintains a
set of private local data $\mathcal{D}_{i}$ and has a local FL runtime
to train a \emph{local model} $\mathbf{w}_{i}$. We use $\mathcal{W}\subseteq\mathcal{N}$
to denote the set of $W$ FL workers selected from data owners. Different
from the traditional centralized training that collects all local
data $\mathcal{D}_{\mathcal{W}}=\cup_{i\in\mathcal{W}}\mathcal{D}_{i}$,
the FL platform only collects and aggregates the updated local models
$\cup_{i\in\mathcal{W}}\mathbf{w}_{i}$ from workers to generate a
\emph{global model} $\mathbf{w}_{g}$. We assume that the data owners
are honest to use their real private data to do training and submit
the true local models to the platform. The FL training process generally
contains the following $3$ steps, where Steps $2$ and $3$ form
an iterative loop between the platform and the workers.
\begin{itemize}
\item Step 1 (task initialization): The platform determines the training
task, i.e., the target application, and the corresponding data requirements.
Meanwhile, it specifies the hyper-parameters of the machine learning
model and the training process. Then, the platform transmits the task
information and the initial global model $\mathbf{w}_{g}^{0}$ to
all workers. 
\item Step 2 (local model training and update): Based on the global model
$\mathbf{w}_{g}^{k}$, where $k$ denotes the current global epoch
index, each worker respectively uses the local data and device to
update the local model parameters $\mathbf{w}_{i}^{k}$. The worker
$i$'s goal in epoch \emph{$k$} is to make parameters $\mathbf{w}_{i}^{k}$
that minimize the predefined loss function $L(\mathbf{w}_{i}^{k})$,
i.e., 
\begin{equation}
\mathbf{w}_{i}^{k^{*}}=\arg\min_{\mathbf{w}_{i}^{k}}L(\mathbf{w}_{i}^{k}).\label{eq:local_training_goal}
\end{equation}
\item Step 3 (global model aggregation and update): The platform receives
and aggregates the local models from workers, and then sends the updated
global model parameters $\mathbf{w}_{g}^{k+1}$ back. The platform
aims to minimize the global loss function $L(\mathbf{w}_{g}^{k})$,
i.e., 
\[
L(\mathbf{w}_{g}^{k})=\frac{1}{W}\sum_{i\in\mathcal{W}}L(\mathbf{w}_{i}^{k}).
\]
\end{itemize}
Steps $2$-$3$ repeat until the global loss converges. Note that
the federated training process can be adopted for various machine
learning approaches based on the gradient descent method such as Support
Vector Machines (SVM), convolutional neural network, and linear regression.
The worker $i$'s local training dataset $\mathcal{D}_{i}$ usually
contains a set of $n_{i}$ feature vectors $\mathbf{x}_{i}=\{x_{1},\ldots,x_{n_{i}}\}$
and a set of corresponding labels $\mathbf{y}=\{y_{1},\ldots,y_{n_{i}}\}$.
Let $\hat{y_{j}}=f(x_{j};\mathbf{w})$ denote the predicted result
from the model $\mathbf{w}$ using data vector $x_{j}$. We focus
on the neural network model in which a common loss function is the
mean square error (MSE) defined as 
\begin{equation}
l(\mathbf{w}_{i}^{k})=\frac{1}{n_{i}}\sum_{j=1}^{n_{i}}(y_{j}-f(x_{j};\mathbf{w}_{i}^{k}))^{2}.\label{eq:mse_loss}
\end{equation}

Global model aggregation is the core part of the FL scheme. In this
paper, we apply the classical federated averaging algorithm (FedAvg)~\cite{BrendanMcMahan2016}
in Algorithm~\ref{alg:AveragingAlg}. According to (\ref{eq:local_training_goal}),
the worker $i$ trains the local model on minibatches sampled from
the original local dataset (lines $4$-$8$). At the $k$th iteration,
the platform minimizes the global loss using the averaging aggregation
which is formally defined as 
\begin{equation}
\mathbf{w}_{g}^{k}=\frac{1}{\sum_{i\in\mathcal{W}}n_{i}}\sum_{i\in\mathcal{W}}n_{i}\mathbf{w}_{i}^{k}.\label{eq:averaging_aggr}
\end{equation}
\begin{algorithm}[tbh]
\footnotesize
\begin{algorithmic}[1]
\Require{Local minibatch size $\delta_{B}$, number of local epochs $\delta_{l}$, number of global epochs $\delta_{g}$, and learning rate $\eta$.}
\Ensure{Global model $\mathbf{w}_{g}$.}
\State{[Worker $i$]}
\State{\textbf{LocalTraining}($i$, $\mathbf{w}_i$):}
	\State{Split the local dataset $\mathcal{D}_i$ to minibatches and include them into the set $\mathcal{B}_i$.}
	\For{each local epoch from $1$ to $\delta_{l}$}\qquad (stochastic gradient descent (SGD))
		\For{each minibatch in $\mathcal{B}_i$ }
			\State{$\mathbf{w} \gets \mathbf{w} - \eta l'(\mathbf{w})$ \qquad ($l'$ is the gradient of $l$ on the minibatch.)}
		\EndFor
	\EndFor
	\State{}
	\State{[Platform]}
	\State{Initialize $\mathbf{w}_{g}^{0}$}
	\For{each global epoch $k$ from $1$ to $\delta_{g}$}
		\State{Randomly choose a subset of $\delta_{s}$ workers from $\mathcal{W}$}
		\For{each worker $i$ in the sampled subset $\textbf{parallely}$}
			\State{$\mathbf{w}_{i}^{k+1} \gets \textbf{LocalTraining}$($i$, $\mathbf{w}_{g}^{k}$)}
		\EndFor
\State{$\mathbf{w}_{g}^{k}=\frac{1}{\sum_{i\in\mathcal{W}}n_{i}}\sum_{i\in\mathcal{W}}n_{i}\mathbf{w}_{i}^{k}$ \qquad (Averaging aggregation)} 
	\EndFor

\end{algorithmic} 

\caption{Federated averaging algorithm (FedAvg)~\cite{BrendanMcMahan2016}\label{alg:AveragingAlg}}
\end{algorithm}
As the hyper-parameters of Algorithm~\ref{alg:AveragingAlg}, $\delta_{B}$ is the local minibatch size, $\delta_{\mathrm{l}}$ is the number of local epochs and $\delta_{\mathrm{g}}$ is the number of global epochs and $\eta$ is the learning rate.  

\subsection{Local Data Evaluation\label{subsec:Local_Data_Evaluation}}

The evaluation of local data is the first step for both the data owners and the platform in the valuation of FL service. The data owner needs to calculate the cost of collecting the local data. The local data cost not only comes from the deployment of sensing devices, e.g., IoT gadgets and smart phones, but also from the data pre-processing that requires costly human intervention for data annotation and cleaning, e.g., redundancy elimination and anomaly detection. Hence, the data owner $i$ has a unit cost $\gamma_{i}>0$ of local data. The local data cost $c_{i}^{\mathrm{d}}$ can be written as  
\begin{equation}
c_{i}^{\mathrm{d}}=d_{i}\gamma_{i}\label{eq:worker_data_cost}
\end{equation}
where $d_{i}>0$ is the data owner $i$'s local data size.

The platform cares about the data quality and needs a metric to quantify data owners' potential contributions to the task completion. Due to the unique features of local data in FL, we focus on two critical attributes of local data: one is the \emph{data size} and the other one is the \emph{data distribution}. According to~\cite{Domingos2012} and the experimental validation in~\cite{Jiao2018}, data size plays an essential role in improving the data quality where more data generally means better prediction performance. With respect to the data distribution, the conventional centralized learning, e.g., data center learning, usually assumes that the training data are independently and identically distributed (IID). However, the local data are user-specific and usually non-IID in the FL scenario. The characteristic of non-IID dominantly affects the performance,
e.g., prediction accuracy, of the trained FL model~\cite{BrendanMcMahan2016}.
Indicated in~\cite{Zhao2018}, the accuracy reduction is mainly
due to the weights divergence which can be quantized by the earth
mover\textquoteright s distance (EMD) metric. A large EMD value means
that the weights divergence is high which adversely affects the global
model quality. We consider an $L$ class classification problem defined
over a compact space $\mathcal{X}$ and a label space $\mathcal{Y}$.
The data owner $i$'s data samples $\mathcal{D}_{i}=\left\{ \mathbf{x}_{i},\mathbf{y}_{i}\right\} $
distribute over $\mathcal{X}\times\mathcal{Y}$ following the distribution
$\mathbb{P}_{i}$. Let $\sigma_{i}$ denote the EMD of $\mathcal{D}_{i}$.
Specifically, given the actual distribution $\mathbb{P}_{a}$ for
the whole population, the EMD $\sigma_{i}$ is calculated by~\cite{Zhao2018}
\begin{equation}
\sigma_{i}=\sum_{j\in\mathcal{Y}}\left\Vert \mathbb{P}_{i}(y=j)-\mathbb{P}_{a}(y=j)\right\Vert .\label{eq:emd_definition}
\end{equation}
The actual distribution $\mathbb{P}_{a}$ is actually used as a reference
distribution. It can be the public knowledge or announced by the platform
which has sufficient historical data to estimate $\mathbb{P}_{a}$.

Let $\boldsymbol{\sigma}=\{\sigma_{1},\ldots,\sigma_{N}\}$ denote
the set of all data owner's EMD value. With the data size and the
EMD metric, the FL platform can measure its data utility. The real-world
experimental results in Section~\ref{sec:Experimental-and-simulation}
indicate that the relationship between the \emph{model quality} $q$,
e.g., prediction accuracy, and the selected workers' total data size
$D$ and average EMD $\Delta$ can be well represented by the following
function:
\begin{align}
q(\mathcal{W}) & =q(D(\mathcal{W}),\Delta(\mathcal{W}))\nonumber \\
 & =\alpha(\Delta)-\kappa_{1}\mathrm{e}^{-\kappa_{2}(\kappa_{3}D)^{\alpha(\Delta)}}\label{eq:quality_func}
\end{align}
where $D$ and $\Delta$ are functions of the set of workers $\mathcal{W}$,
i.e., the total data size $D(\mathcal{W})=\sum_{i\in\mathcal{W}}d_{i}$
and the average EMD metric $\Delta(\mathcal{W})=\frac{\sum_{i\in\mathcal{W}}\sigma_{i}}{\left|\mathcal{W}\right|}$
with $\Delta(\emptyset)=0$, and $\alpha(\Delta)=\kappa_{4}\exp(-(\frac{\Delta+\kappa_{5}}{\kappa_{6}})^{2})<1$.
$\kappa_{1},\ldots,\kappa_{6}>0$ are positive curve fitting parameters. The curve fitting approach for determining the function of machine learning quality is typical in the literature and a similar function has been adopted in other works, such as \cite{Liu2018}. In the experiment presented in Section~\ref{subsec:Veri-for-Data-util}, the data utility
function (\ref{eq:quality_func}) fits well when $\sigma$ falls in the $[0,\sigma_{\max}]$. To guarantee good service quality, $\sigma_{\max}$ can be set as the maximum EMD that the platform can accept. The first term $\alpha(\Delta)$ reflects that the increasing average EMD metric causes the degradation of the model performance. The exponential term $-\kappa_{1}\mathrm{e}^{-\kappa_{2}(\kappa_{3}D)^{\alpha(\Delta)}}$ captures the diminishing marginal returns when the total data size increases. Hereby, we define the platform's data utility $\varphi$ as a linear function of $q$ as 
\begin{align}
\varphi(\mathcal{W}) & =\varphi(D(\mathcal{W}),\Delta(\mathcal{W}))\nonumber \\
 & =\kappa_{7}q(\mathcal{W})\nonumber \\
 & =\kappa_{7}\left(\alpha(\Delta)-\kappa_{1}\mathrm{e}^{-\kappa_{2}(\kappa_{3}D)^{\alpha(\Delta)}}\right)
\end{align}
where $\kappa_{7}$ represents the profit per unit performance. 

\subsection{Auction based FL Services Market\label{subsec:Reverse-Auction-Mechanism}}

To recruit enough qualified workers for successful federated training, the FL platform\footnote{We use ``FL platform'' and ``platform'' interchangeably.} conducts an auction. Figure~\ref{fig:FL_market} depicts the auction supported the FL process. For simplicity, we assume that the data owners' computing and storage capabilities, i.e., the CPU frequency and memory, can meet the FL platform's minimum requirement of the
training speed and the local model size. Since the communication delay seriously degrades the efficiency of FL~\cite{Kairouz2019}, the platform requires the FL worker\footnote{Note that the FL worker refers to as the data owner that has been selected by the platform to perform the FL training.} to immediately transmit back the updated local model at the transmission rate $R$ bits/s when the local training is completed. With the fixed model size, this is actually equivalent to requiring the workers to finish the model transmission in a fixed time.

As described in Step $1$ in Section~\ref{subsec:Preliminaries_of_FL}, the platform first
initializes the global neural network model with size $M$ and hyper-parameters, such as $\delta_{l},\delta_{g}$ and $R$. Then, the platform announces the auction rule and advertises the FL task to the data owners. Then, the data owners report their type profile $\mathbf{T}=\{\mathbf{t}_{1},\ldots,\mathbf{t}_{N}\}$ and the requested wireless channel profile $\mathbf{\mathcal{C}}=\{\mathcal{C}_{1},\ldots,\mathcal{C}_{N}\}$. The data owner $i$'s type $\mathbf{t}_{i}$ contains the bid $b_{i}$ which reveals its private service cost/valuation $c_{i}$, the size $d_{i}$ and EMD value $\sigma_{i}$ of its possessed local data, i.e., $\mathbf{t}_{i}=\{ b_{i},d_{i},\sigma_{i}\}$.  $\mathcal{C}_{i}$ is the set of data owner $i$'s requested wireless channels to communicate with the FL platform.

Since this paper focuses on the resource allocation of the FL system, we assume that there is no adverse attack in the FL training. The data owners cannot provide services with higher data quality than their truly owned. They will not report higher data size or lower EMD metric to the platform. Otherwise, this would be seen as the \emph{model update poisoning attack}~\cite{Kairouz2019}. Based on the received types, the platform has to select workers and notifies all data owners the service allocation, i.e., the set of FL workers $\mathcal{W}$, and the corresponding payments $\mathbf{p}=\left\{ p_{1},\ldots,p_{N}\right\} $
to each data owner. The workers are considered to be \emph{single-minded}
at the channel allocation. That is, the data owner $i$ only accepts the set of its requested channels if it wins the auction. The payment for a data owner failing the auction is set to be zero, i.e., $p_{i}=0$ if $i\notin\mathcal{W}$. Once the auction results are released,
an FL session starts and the selected workers train the local model
using their own local data. Meanwhile, the platform keeps aggregating the
local models and updating the global model. Finally, the platform
pays the workers when the FL session is completed. 

\subsection{Service Cost in the FL Market\label{subsec:Service-Cost}}

Besides the local data cost defined in~(\ref{eq:worker_data_cost}), the
data owner also needs to calculate the costs of computation and communication
to estimate its service cost if it becomes the worker. According to our previous experimental results about the energy consumption of the FL training~\cite[Figure 2]{Zou2019}, the data owner $i$'s computational cost $c_{i}^{\mathrm{p}}$ is defined as a linear function of the data size $d_i$, which is written as 
\begin{equation}
c_{i}^{\mathrm{p}}=d_{i}\delta_{\mathrm{l}}\delta_{\mathrm{g}}M\alpha_{i}\label{eq:worker_computation_cost}
\end{equation}
where $\alpha_{i}$ is the data owner $i$'s unit computational cost.
Since the structures of the global model and the local model are the
same when applying FedAvg, we use $M$ to denote the model size. With
respect to the communication cost, we ignore the communication overhead
and assume the channel is slow-fading and stable. Since this paper
focuses on the design of incentive mechanism, we consider a frequency-division
multiple-access (FDMA) communication scheme. This is also for simplicity
and minimum communication interference. Nonetheless, other more sophisticated
wireless communication configurations can be adopted with slight modification
in the cost function. 

According to Shannon's formula~\cite{Shannon1948}, the data owner $i$'s communication power cost is
\begin{align}
P_{i}^{\mathrm{m}}=\frac{(2^{\frac{R}{BC_{i}}}-1)BC_{i}}{h_{i}},
\end{align}
where $B$ is the channel bandwidth, $C_{i}=\left|\mathcal{C}{}_{i}\right|$ is the number of data owner $i$'s requested channels, $BC_{i}$ is the total bandwidth, $h_{i}=\frac{\tilde{h}_{i}^{2}}{\psi_{0}}$ is the normalized channel power gain, $\tilde{h}_{i}$ is the channel gain between the data owner $i$ and the FL platform (as a base station), and $\psi_{0}$ is the one-sided noise power spectral density. The total cost for communication is 
\begin{align}
c_{i}^{\mathrm{m}} & =P_{i}^{\mathrm{m}}\frac{M}{R}\delta_{\mathrm{g}}\beta_{i}\label{eq:worker_communication_cost-1}\\
 & =\frac{(2^{\frac{R}{BC_{i}}}-1)BC_{i}M\delta_{\mathrm{g}}\beta_{i}}{h_{i}R}\label{eq:worker_communication_cost}
\end{align}
where $\frac{M}{R}\delta_{\mathrm{g}}$ is the total time for model transmission, $\beta_{i}$ is the data owner $i$'s unit energy cost for communication. The channel conditions of different subcarriers for each data owner can be perfectly estimated. That is, $h_{i}$
is known by both the data owner $i$ and the platform. Adding all costs in (\ref{eq:worker_data_cost}), (\ref{eq:worker_computation_cost}) and (\ref{eq:worker_communication_cost-1}) together, the data owner $i$'s total service cost $c_{i}$ is 
\begin{align}
c_{i} & =c_{i}^{\mathrm{d}}+c_{i}^{\mathrm{p}}+c_{i}^{\mathrm{m}}\nonumber \\
 & =d_{i}\gamma_{i}+d_{i}\delta_{\mathrm{l}}\delta_{\mathrm{g}}M\alpha_{i}+\frac{(2^{\frac{R}{BC_{i}}}-1)BC_{i}M}{h_{i}R}\delta_{\mathrm{g}}\beta_{i}.\label{eq:worker_cost}
\end{align}
Since our proposed auction mechanisms are truthful (to be proved later), the reported bid $b_{i}$ is equal to the true service cost $c_{i}$, i.e., $b_{i}=c_{i}$. 

Similarly, the FL platform has the computational cost $\hat{c}^{\mathrm{p}}$ for model averaging and the communication cost $\hat{c}^{\mathrm{m}}$ for global model transmission defined as follows:
\begin{equation}
\hat{c}^{\mathrm{p}}(\mathcal{W})=\delta_{\mathrm{g}}M(W-1)\hat{\alpha},\label{eq:platform_computation_cost}
\end{equation}
\begin{equation}
\hat{c}^{\mathrm{m}}(\mathcal{W})=\sum_{i\in\mathcal{W}}\frac{(2^{\frac{R}{BC_{i}}}-1)BC_{i}M}{h_{i}R}\delta_{\mathrm{g}}\hat{\beta},\label{eq:platform_communication_cost}
\end{equation}
where $\hat{\alpha}$ and $\hat{\beta}$ are respectively the unit costs for computation and communication. Hence, we have the platform's total cost as follows
\begin{align}
\hat{c}(\mathcal{W}) & =\hat{c}^{\mathrm{p}}+\hat{c}^{\mathrm{m}}\\
 & =\delta_{\mathrm{g}}M(W-1)\hat{\alpha}+\sum_{i\in\mathcal{W}}\frac{(2^{\frac{R}{BC_{i}}}-1)BC_{i}M}{h_{i}R}\delta_{\mathrm{g}}\hat{\beta}.\label{eq:platform_total_cost}
\end{align}

\subsection{Social Welfare Optimization and Desired Economic Properties\label{subsec:SW-Opti-Econo}}

With the data utility and the service cost introduced in Sections~\ref{subsec:Local_Data_Evaluation} and~\ref{subsec:Service-Cost}, we can obtain the utility functions of all entities. The FL platform's utility is the data utility minus
the total cost and the total payments to workers, which is written as
\begin{align}
\hat{u} & =\varphi(D,\Delta)-\hat{c}-\sum_{i\in\mathcal{W}}p_{i}.\label{eq:platform_utility_origin}
\end{align}
The data owner $i$'s utility is the difference between its payment
$p_{i}$ and service cost $c_{i}$, which is expressed as 
\begin{equation}
u_{i}=p_{i}-c_{i}.\label{eq:worker_i_utility}
\end{equation}

In Section~\ref{sec:Reverse_multidimenstional_auction}, we design
the auction mechanism to maximize the \emph{social welfare} which
can be regarded as the FL system efficiency~\cite{Zhang2017} and
is defined as the sum of the platform's utility and the data owners'
utilities. Formally, the social welfare maximization problem is
\begin{align}
\max_{\mathcal{W}\subseteq\mathcal{N}}S(\mathcal{W}) & =\hat{u}+\sum_{i\in\mathcal{W}}u_{i}\label{eq:sw_maixmization}\\
 & =\varphi(D(\mathcal{W}),\Delta(\mathcal{W}))-\hat{c}(\mathcal{W})-\sum_{i\in\mathcal{W}}c_{i}\\
\mathrm{s.t.} & \:\:\:\mathcal{C}{}_{i}\cap\mathcal{C}{}_{j}=\emptyset,\forall i,j\in\mathcal{W},i\neq j.\label{eq:FDD_channels_constraint}
\end{align}
As we consider the FDD communication scheme, the constraint in (\ref{eq:FDD_channels_constraint})
requires that the sets of workers allocated channels
have no conflict with each other. For an efficient and stable FL market,
the following economic properties should be guaranteed.
\begin{itemize}
\item \emph{Truthfulness (Incentive compatibility, IC).} The data owner
$i$ has no incentive to report a fake type for a higher utility.
Formally, with other data owner's types fixed, the condition for the
truthfulness is 
\[
u_{i}(t'_{i})\leq u_{i}(t{}_{i}),\forall t'_{i}\neq t_{i},
\]
where $t_{i}=(b_{i},d_{i},\sigma_{i})$ is data
owner $i$'s true type and $t'_{i}=(b'_{i},d'_{i},\sigma'_{i})$
is a false type. 
\item \emph{Individual rationality (IR).} No data owner will suffer a deficit
from its FL service provision, i.e., $u_{i}(t{}_{i})\geq0$,$\forall i\in\mathcal{N}$.
\item \emph{Computational efficiency (CE).} The auction algorithm can be
completed in polynomial time. 
\end{itemize}

\section{Reverse Multi-dimensional Auction Mechanism for Federated Training \label{sec:Reverse_multidimenstional_auction}}

In this section, we first design a truthful auction mechanism, called Reverse Multi-dimensional auction (RMA) mechanism, to maximize the social welfare defined in (\ref{eq:sw_maixmization}). As presented in Algorithm~\ref{alg:CRM}, the RMA generally follows a randomized and greedy way to choose the FL workers and decides the payments.
It consists of three consecutive phases: dividing (lines 2-9), worker
selection (lines 12-20) and service payment determination (lines 21-41). 

The RMA first divides the workers into $G$ groups, i.e., $\left\{ \Theta_{1},\ldots,\Theta_{j},\ldots,\Theta_{G}\right\} $,
according to the EMD metric. Each group consecutively covers an EMD
interval $\epsilon=\frac{\sigma_{\max}}{G}$. That is, the data owner
$i$ whose EMD value $\sigma_{i}$ falls in $[(j-1)\frac{\sigma_{\max}}{G},j\frac{\sigma_{\max}}{G})$
will be put in the group $\Theta_{j}$. Meanwhile, we define a \emph{virtual
EMD} \emph{value} for the data owner\emph{ $i$} in group $j$ by
the corresponding interval midpoint, i.e., $\widetilde{\sigma}^{j}=\frac{(2j-1)\sigma_{\max}}{2G}$.
For group $j$, the virtual social welfare $\widetilde{S}^{j}(\mathcal{W})$
is calculated by using the virtual EMD value as follows:
\begin{align}
\widetilde{S}^{j}(\mathcal{W}) & =\widetilde{\varphi}^{j}(\mathcal{W})-\hat{c}(\mathcal{W})-\sum_{i\in\mathcal{W}}b_{i}\label{eq:virtual_SW-1}\\
 & =\varphi(D(\mathcal{W}),\widetilde{\Delta}^{j}(\mathcal{W}))-\hat{c}(\mathcal{W})-\sum_{i\in\mathcal{W}}b_{i}
\end{align}
where $\widetilde{\varphi}^{j}(\mathcal{W})=\varphi(D(\mathcal{W}),\widetilde{\Delta}^{j}(\mathcal{W}))$
and $\widetilde{\Delta}^{j}(\mathcal{W})=\frac{\sum_{i\in\mathcal{W}}\widetilde{\sigma}^{j}}{\left|\mathcal{W}\right|}=\widetilde{\sigma}^{j}=\frac{(2j-1)\sigma_{\max}}{2G}$.
Let $\mathcal{L}(\mathcal{W})$ denote the set of workers that have
channel conflicts with the worker set $\mathcal{W}$. We introduce
the \emph{marginal virtual social welfare density} $V_{i}^{j}(\mathcal{W})$
for the worker $i$ in group $j$ defined as
\begin{align}
V_{i}^{j}(\mathcal{W}) & =\frac{\widetilde{S}^{j}(\mathcal{W}\cup\{i\})-\widetilde{S}^{j}(\mathcal{W})}{\left|\mathcal{L}(\{i\})\right|}\label{eq:marginal_density_virtual_SW}\\
 & =\frac{1}{\left|\mathcal{L}(\{i\})\right|}\left(\kappa_{1}\kappa_{7}\mathrm{e}^{-\kappa_{2}(\kappa_{3}\sum_{k\in\mathcal{W}}d_{k})^{\alpha(\widetilde{\Delta}^{j})}}\right.\nonumber \\
 & \left.\;-\kappa_{1}\kappa_{7}\mathrm{e}^{-\kappa_{2}(\kappa_{3}\sum_{k\in\mathcal{W}\cup\{i\}}d_{k})^{\alpha(\widetilde{\Delta}^{j})}}-\hat{c}(\{i\})-b_{i}\right).
\end{align}
 For the sake of brevity, we simply call it \emph{marginal density}.
\begin{algorithm}[tbh]
{\scriptsize{}\footnotesize
\begin{algorithmic}[1] 
\Require{$G$, $\bar{E}$ and $\mathbf{t}=\{t_1,\ldots,t_i, \ldots, t_N\}$ with $t_i=\{b_{i},d_{i},e_{i},\mathcal{C}_{i}\}$.} 
\Ensure{The set of FL workers $\mathcal{W}$ and the service payment~$\mathbf{p}$.} 
\Begin 
	\State{$\epsilon \gets \frac{\sigma_{\max}}{G}, \mathcal{U} \gets \emptyset$, $\mathcal{W}_o \gets \emptyset$, $\mathcal{G} \gets \emptyset$}
	\For{$j=1$ to $G$}
		\State{$\Theta_{j} \gets \emptyset$, $\mathcal{W}_j \gets \emptyset$, $\mathcal{G} \gets \mathcal{G} \cup \{j\}$}
	\EndFor
	\ForEach{$i \in \mathcal{N}$}
		\State{$p_i \gets 0, j \gets \left\lceil e_i/{\epsilon}\right\rceil$}
		\State{$\bar{\gamma}_{i} \gets \frac{(2j-1)\epsilon}{2}$}
	\EndFor
	\While{$\mathcal{G} \neq \emptyset$ }
		\State{Uniformly select $j$ from $\mathcal{G}$}
		\State{$\mathcal{G} \gets  \mathcal{G}\setminus \{j\}$, $\tilde{\Theta}_{j} \gets \Theta_{j}\setminus\mathcal{L}(\mathcal{W}_o)$, $\mathcal{U} \gets \emptyset$}
		\While{$\tilde{\Theta}_{j} \neq \emptyset$}
			\State{$k^* \gets \arg\max_{k\in \tilde{\Theta}_{j}}V_{k}^{j}(\mathcal{U}\cup\mathcal{W}_o)$}
			\If{$V_{k^*}^{j}(\mathcal{U}\cup\mathcal{W}_o)<0$}
				\Break
			\EndIf
			\State{$\mathcal{U} \gets \mathcal{U}\cup\{k^*\}$, $\tilde{\Theta}_{j} \gets \tilde{\Theta}_{j}\setminus(\mathcal{L}(\{k^*\}\cup\{k^*\})$}
			\State{$\mathcal{W}_j \gets \mathcal{U}$}
		\EndWhile

		\ForEach{$i\in \mathcal{W}_j$}
			\State{$\Theta^{-i}_{j} \gets  \Theta_{j}\setminus (\{i\}\cup \mathcal{L}(\mathcal{W}_o))$, $\tilde{\Theta}^{-i}_{j}\gets\Theta^{-i}_{j}$, $\mathcal{T} \gets \emptyset$}
			\If{$\tilde{\Theta}^{-i}_{j} = \emptyset$}
				\State{$p_i \gets \arg_{b_{i}}V_{i}^{j}(\mathcal{T}\cup\mathcal{W}_o)=0$}
			\EndIf
			\While{$\tilde{\Theta}^{-i}_{j} \neq \emptyset$}
				\State{$i_{k^*} \gets \arg\max_{i_k\in\tilde{\Theta}^{-i}_{j}}$ $V_{i_k}^{j}(\mathcal{T}\cup\mathcal{W}_o)$}

				\If{$V_{i_{k^*}}^{j}(\mathcal{T}\cup\mathcal{W}_o)<0$}
					\State{$p_i \gets \max\{p_i, \arg_{b_{i}}V_{i}^{j}(\mathcal{T}\cup\mathcal{W}_o)=0\}$}
					\Break
				\ElsIf{$i \in \mathcal{L}(\{i_{k^*}\})$}
					\State{$p_i \gets \arg_{b_{i}}V_{i}^{j}(\mathcal{T}\cup\mathcal{W}_o)=V_{i_{k^*}}^{j}(\mathcal{T}\cup\mathcal{W}_o)$}
					\Break
				\EndIf
				\State{$p_i \gets \max\{p_i, \arg_{b_{i}}V_{i}^{j}(\mathcal{T}\cup\mathcal{W}_o)=V_{i_{k^*}}^{j}(\mathcal{T}\cup\mathcal{W}_o)\}$}
				\State{$\mathcal{T} \gets \mathcal{T}\cup\{i_{k^*}\}$, $\tilde{\Theta}^{-i}_{j} \gets \tilde{\Theta}^{-i}_{j} \setminus (\{i_{k^*}\}\cup\mathcal{L}(\{i_{k^*}\}))$}
				\If{$\tilde{\Theta}^{-i}_{j}=\emptyset$}
					\State{$p_i \gets \max\{p_i, \arg_{b_{i}}V_{i}^{j}(\mathcal{T}\cup\mathcal{W}_o)=0\}$}
				\EndIf

			\EndWhile
		\EndFor
		\State{$\mathcal{W}_o \gets \mathcal{W}_o\cup{\mathcal{W}_j}$}
	\EndWhile
	\State{$\mathcal{W} \gets \mathcal{W}_o$}
\End
\end{algorithmic}\caption{Reverse Multi-dimensional auction (RMA)\label{alg:CRM}}
}
\end{algorithm}
We use $\mathcal{W}_{o}$ to denote the set of already selected workers
from other groups. In each group $j$, the RMA first excludes the
workers that are conflicted with $\mathcal{W}_{o}$, i.e., $\tilde{\Theta}_{j}=\Theta_{j}\setminus\mathcal{L}(\mathcal{W}_{o})$.
Then, the RMA finds and sorts the data owners which have no channel
conflict with each other in $\tilde{\Theta}_{j}$ by non-increasing
order of the marginal density: 
\begin{align}
V_{1}^{j}(\mathcal{U}_{0}\cup\mathcal{W}_{o}) & \geq V_{2}^{j}(\mathcal{U}_{1}\cup\mathcal{W}_{o})\geq\cdots\nonumber \\
\geq V_{k}^{j}( & \mathcal{U}_{k-1}\cup\mathcal{W}_{o})\geq\cdots\geq V_{K'}^{j}(\mathcal{U}_{K'-1}\cup\mathcal{W}_{o})\label{eq:winner_sorting}
\end{align}
where $\mathcal{U}_{k-1}$ is the set of first $k-1$ sorted data
owners and $\mathcal{U}_{0}=\emptyset$. There are totally $K'$ data
owners in the sorting and the $k$th data owner has the largest marginal
density $V_{k}^{j}(\mathcal{U}_{k-1}\cup\mathcal{W}_{o})$ in $\tilde{\Theta}_{j}\setminus\mathcal{U}_{k-1}$
while having no channel conflict with data owners in $\mathcal{U}_{k-1}$.
From the sorting, the RMA aims to find the set $\mathcal{U}_{K_{s}}$
containing $K_{s}$ data owners as workers, such that $V_{K_{s}}^{j}(\mathcal{U}_{K_{s}-1}\cup\mathcal{W}_{o})>0$
and $V_{K_{s}+1}^{j}(\mathcal{U}_{K_{s}}\cup\mathcal{W}_{o})$ (lines
12-19).

Once the set of  workers in group $j$ has been determined, the RMA
re-executes the worker selection on the set of data owners in group
$j$ (except the data owner $i$), i.e., $\tilde{\Theta}_{j}^{-i}=\tilde{\Theta}_{j}\setminus\left\{ i\right\} $,
to calculate the payment $p_{i}$ for worker $i$ (lines 22-34). Similarly,
the RMA sort data owners in $\tilde{\Theta}_{j}^{-i}=\tilde{\Theta}_{j}\setminus\left\{ i\right\} $
as follows: 
\begin{align}
V_{i_{1}}^{j}(\mathcal{T}_{0}\cup\mathcal{W}_{o}) & \geq V_{i_{2}}^{j}(\mathcal{T}_{1}\cup\mathcal{W}_{o})\geq\cdots\nonumber \\
\geq V_{i_{k}}^{j}(\mathcal{T}_{k-1} & \cup\mathcal{W}_{o})\geq\cdots\geq V_{i_{K''}}^{j}(\mathcal{T}_{K''-1}\cup\mathcal{W}_{o})\label{eq:payment_sorting}
\end{align}
where $\mathcal{T}_{k-1}$ is the set of the first $k-1$  data owners
in the sorting and $\mathcal{T}_{0}=\emptyset$. From the sorting,
we select the first $K_{p}$ data owners as the workers where the
$K_{p}$th data owner $i_{K_{p}}$ is (1) the first one that has a
non-negative marginal density and channel conflicts with worker $i$,
i.e., $i\in\mathcal{L}(\{i_{K_{p}}\})$ and $V_{i_{K_{p}}}^{j}(\mathcal{T}_{i_{K_{p}-1}}\cup\mathcal{W}_{o})\geq0$,
\emph{or} (2) the last one that satisfies $i\notin\mathcal{L}(\{i_{k}\})$
and $V_{i_{K_{p}}}^{j}(\mathcal{T}_{i_{K_{p}-1}}\cup\mathcal{W}_{o})\geq0$.
If the data owner $i_{K_{p}}$ is chosen by the condition (1), the
payment $p_{i}$ is set to be the bid value such that the worker $i$
and the data owner $i_{K_{p}}$ have equal marginal density on $\mathcal{T}_{i_{K_{p}-1}}\cup\mathcal{W}_{o}$,
i.e., $p_{i}\gets\arg_{b_{i}}V_{i}^{j}(\mathcal{T}_{i_{K_{p}-1}}\cup\mathcal{W}_{o})=V_{i_{K_{p}}}^{j}(\mathcal{T}_{i_{K_{p}-1}}\cup\mathcal{W}_{o})$
(lines 31-33). If data owner $i_{K_{p}}$ is chosen by condition (2),
$p_{i}$ is set to be the maximum value such that $V_{i}^{j}(\mathcal{T}_{i_{k-1}}\cup\mathcal{W}_{o})\geq V_{i_{k}}^{j}(\mathcal{T}_{i_{k-1}}\cup\mathcal{W}_{o})$,
$\exists k\in\{1,\ldots,K_{p}\}$ or $V_{i}^{j}(\mathcal{T}_{K_{p}}\cup\mathcal{W}_{o})\geq0$
(lines 28-30 and 35-39). 

The dividing phase decomposes the original auction mechanism $\mathbf{M}_{o}$
into a set of $G$ sub-auctions. We use $\mathbf{M}_{j\in\{1,\ldots,G\}}$
to denote the sub-auction mechanism for group $j$. Since the data
owners in each group have the same EMD value and the reported channel
information is true, only the bid and the data size $(b_{i},d_{i})$
in the type $t_{i}$ can be manipulated. Thus, each sub-auction can
be reduced to a deterministic reverse multi-unit auction where each
data owner $i$ bids $b_{i}$ to sell $d_{i}$ data units. Reflected
in the data utility function in~(\ref{eq:quality_func}), the $d_{i}$
data units here essentially represent the data owner $i$'s service
quality. Here, again, the data owners are \emph{single-minded}, which
means they can only sell the reported amount of data units. The deterministic
auction mechanism here means the same input types will deterministically
generate the same unique output. As the randomization is applied over
a collection of deterministic mechanisms (line 11), the original auction
mechanism $\mathbf{M}_{o}$ is a \emph{randomized auction mechanism}~\cite{Archer2004}.
Our design rationale of each sub-auction is formally presented in
Theorem~\ref{thm:multi_unit_charac} which adopts the characterizations
for the truthful forward multi-unit auction presented in~\cite[Section 9.5.4]{Nisan2015}.

\begin{thm}
\label{thm:multi_unit_charac}In the reverse multi-unit and single-minded
setting, an auction mechanism is truthful if it satisfies the following
two properties:
\end{thm}
\begin{enumerate}
\item Monotonicity: If a bidder $i$ wins with type $(b_{i},d_{i})$, then
it will also win with any type which offers at most as much price
for at least as many items. That is, bidder $i$ will still win if
the other bidders do not change their types and bidder $i$ changes
its type to some $(b_{i}',d_{i}')$ with $b_{i}\ge b_{i}'$ and $d_{i}\le d'_{i}$.
\item Critical payment: The payment of a winning type $(b_{i},d_{i})$ by
bidder $i$ is the largest value needed in order to sell $d_{i}$
items, i.e., the supremum of $b_{i}'$ such that $(b_{i}',d_{i})$
is still a winning type, when the other bidders do not change their
types.
\end{enumerate}
We next show the desired properties of the RMA, including the truthfulness
(Proposition~\ref{prop:crma-IC}), the individual rationality (Proposition~\ref{prop:CRM-IR})
and the computational efficiency (Proposition~\ref{prop:CRM-EFFICIENT}). 
\begin{prop}
\label{prop:crma-IC}The RMA mechanism is universally truthful (incentive
compatible).
\end{prop}
\begin{IEEEproof}
We first investigate the truthfulness of the sub-auction $M_{j}$.
Since the RMA guarantees that data owners in the same group have the
same virtual EMD value and the group selection is random (line 11),
data owners have no incentive to report false EMD value. Therefore,
we just need to discuss the truthfulness of the reported data size
and the bid. According to Theorem~\ref{thm:multi_unit_charac}, it
suffices to prove that the worker selection of $M_{j}$ is monotone,
and the payment $p_{i}$ is the critical value for the  data owner
$i$ to win the auction. Given a fixed EMD value $\Delta$, we construct
a function $o(z)$ as 
\begin{equation}
o(z)=\kappa_{1}\kappa_{7}\mathrm{e}^{-\kappa_{2}(\kappa_{3}z)^{\alpha(\Delta)}}\label{eq:1st_derivative_sz}
\end{equation}
where $x\in\mathbb{R}^{+}$ and $\alpha(\Delta)\in(0,1)$, $\kappa_{1},\kappa_{2},\kappa_{3},\kappa_{7}\in(0,+\infty)$
are parameters. The first derivative and the second derivative of
$o(z)$ are receptively
\begin{equation}
\frac{\mathrm{d}o(z)}{\mathrm{d}z}=-\kappa_{1}\kappa_{2}\kappa_{3}\kappa_{7}(\kappa_{3}z)^{\alpha(\Delta)-1}\alpha(\Delta)\mathrm{e}^{-\kappa_{2}(\kappa_{3}z)^{\alpha(\Delta)}},
\end{equation}

\begin{align}
\frac{\mathrm{d}^{2}o(z)}{\mathrm{d}z^{2}} & =\kappa_{1}\kappa_{2}\kappa_{7}\alpha(\Delta)\mathrm{e}^{-\kappa_{2}(\kappa_{3}z)^{\alpha(\Delta)}}\nonumber \\
 & \quad\quad(\kappa_{3}z)^{\alpha(\Delta)}(\kappa_{2}\alpha(\Delta)(\kappa_{3}z)^{\alpha(\Delta)}-\alpha(\Delta)+1).
\end{align}
Since $1>\alpha(\Delta)>0$ and $\kappa_{1},\kappa_{2}>0$, we can
find that $\frac{\mathrm{d}o(z)}{\mathrm{d}z}<0$ and $\frac{\mathrm{d}^{2}o(z)}{\mathrm{d}z^{2}}>0$
which means $o(z)$ is a convex and monotonically decreasing function.
Note that expanding $\mathcal{W}$ is equivalent to increasing the
total data size $D(\mathcal{W})=\mathcal{W}=\sum_{k\in\mathcal{W}}d_{k}$.
Substituting $z=\sum_{k\in\mathcal{W}}d_{k}$ and $z=\sum_{k\in\mathcal{W}\cup\{i\}}d_{k}$
into $o(z)$, we can find $V_{i}^{j}(\mathcal{W})=\frac{o(\sum_{k\in\mathcal{W}}d_{k})-o(\sum_{k\in\mathcal{W}\cup\{i\}}d_{k})-\hat{c}(\{i\})-b_{i}}{\left|\mathcal{L}(\{i\})\right|}$
which is monotonically decreasing with $\mathcal{W}$ since $\sum_{k\in\mathcal{W}\cup\{i\}}d_{k}>\sum_{k\in\mathcal{W}}d_{k}$
and the monotonicity and convexity of $o(z)$. It is also clear that
the marginal density $V_{i}^{j}(\mathcal{W})$ defined in (\ref{eq:marginal_density_virtual_SW})
is monotonically decreasing with the bid $b_{i}$ while monotonically
increasing with $d_{i}$. As the data owner $i$ takes the $i$th
place in the sorting (\ref{eq:winner_sorting}), if it changes the
type from $t_{i}$ to $t'_{i}$ by lowering its bid from $b_{i}$
to $b_{i}^{-}$ ($b_{i}>b_{i}^{-}$ ) or raising the reported data
size from $d_{i}$ to $d_{i}^{+}$ ($d_{i}^{+}>d_{i}$ ), it will
have a larger marginal density $V_{i^{'}}^{j}(\mathcal{\mathcal{U}}_{i-1})>V_{i}^{j}(\mathcal{\mathcal{U}}_{i-1})$.
Since $V_{i}^{j}(\mathcal{W})$ is a decreasing function of $\mathcal{W}$,
the data owner $i$'s marginal density can only increase when it is
at a higher rank in the sorting (\ref{eq:winner_sorting}), i.e.,
$V_{i^{'}}^{j}(\mathcal{\mathcal{U}}_{i-k})>V_{i^{'}}^{j}(\mathcal{\mathcal{U}}_{i-1}),\forall k\in\{2,3,\ldots,i\}$.
Thus, we have proved the monotonicity condition required by Theorem~\ref{thm:multi_unit_charac}.

We next prove that $p_{i}$ calculated by Algorithm~\ref{alg:CRM}
is the critical payment, which means that with $d_{i}$ fixed, bidding
a higher price $b_{i}^{+}>p_{i}$ causes the worker $i$ to fail the
auction. As mentioned above, the final payment $p_{i}$ depends on
the data owner $i_{K_{p}}$ in the sorting (\ref{eq:payment_sorting}).
If the $K_{p}$th worker has channel conflict with the worker $i$,
summiting a higher bid $b_{i}^{+}$ makes worker $i$ be ranked after
data owner $i_{K_{p}}$, i.e, $V_{i}^{j}(\mathcal{T}_{K_{p}-1}\cup\mathcal{W}_{o})<V_{i_{K_{p}}}^{j}(\mathcal{T}_{K_{p}-1}\cup\mathcal{W}_{o})$,
and then worker $i$ would be removed from the candidate pool in the
subsequent selection. If the data owner $i_{K_{p}}$ has no channel
conflict with the data owner $i$, a higher bid $b_{i}^{+}>p_{i}$
still causes $V_{i}^{j}(\mathcal{T}_{k-1}\cup\mathcal{W}_{o})<V_{i_{k}}^{j}(\mathcal{T}_{k-1}\cup\mathcal{W}_{o}),\forall k\in\{1,2,\ldots,K_{p}\}$
and $V_{i}^{j}(\mathcal{T}_{K_{p}}\cup\mathcal{W}_{o})<0$, which
apparently cannot lead the data owner $i$ to win the auction. Thus,
the truthfulness of the sub-auction $M_{j}$ is proved. Since each
sub-auction $M_{j}$ is truthful and the original auction mechanism
$\mathbf{M}_{o}$ is a randomization over the collection of the sub-auctions,
we can finally prove that the RMA mechanism is universally truthful~\cite[Definition 9.38]{Nisan2007}.
\end{IEEEproof}
\begin{prop}
\label{prop:CRM-IR}The RMA mechanism is individually rational.
\end{prop}
\begin{IEEEproof}
Let $i_{i}$ denote the worker $i$'s replacement in the payment determination
process, i.e., the $i$th data owner in the sorting (\ref{eq:payment_sorting}).
As the data owner $i_{i}$ must be after the $i$th place in the sorting
(\ref{eq:payment_sorting}) or even not in the sorting if worker $i$
wins the auction, we have $V_{i}^{j}(\mathcal{T}_{i-1}\cup\mathcal{W}_{o})>V_{i_{i}}^{j}(\mathcal{T}_{i-1}\cup\mathcal{W}_{o})$.
As shown in the Algorithm~\ref{alg:CRM}, the payment $p_{i}$ for
worker $i$ is the maximum winning bid $b_{i}^{'}$, which means the
corresponding marginal density $V_{i'}^{j}(\mathcal{T}_{i-1}\cup\mathcal{W}_{o})$
satisfies $V_{i}^{j}(\mathcal{T}_{i-1}\cup\mathcal{W}_{o})>V_{i_{i}}^{j}(\mathcal{T}_{i-1}\cup\mathcal{W}_{o})\geq V_{i'}^{j}(\mathcal{T}_{i-1}\cup\mathcal{W}_{o})$.
Since $V_{i}^{j}(\mathcal{W})$ is monotonically decreasing with the
bid $b_{i}$ (see the proof for Proposition~\ref{prop:crma-IC}),
we have $p_{i}=b'_{i}\geq b_{i}=c_{i}$, which means the worker $i$'s
utility $u_{i}$ defined in (\ref{eq:worker_i_utility}) is non-negative,
i.e., $u_{i}(t{}_{i})\geq0$. Therefore, we can guarantee the individual
rationality of each sub-auction $M_{j}$ and the original RMA mechanism
$M_{o}$. 
\end{IEEEproof}
\begin{prop}
\label{prop:CRM-EFFICIENT}The RMA mechanism is computationally efficient.
\end{prop}
\begin{IEEEproof}
For each sub-auction $M_{j}$ (lines 12-41) in Algorithm~\ref{alg:CRM},
finding the  workers in group $\Theta_{j}$ with the maximum marginal
density has the time complexity of $O(\left|\Theta_{j}\right|)$ (line
14). Since the number of workers is at most $\left|\Theta_{j}\right|$,
the worker selection process (the while-loop lines 13-20) has the
time complexity of $O(\left|\Theta_{j}\right|^{2})$. In the payment
determination process (lines 21-41), each for-loop executes similar
steps as the while-loop in lines 13-20 and the payment determination
process generally has the time complexity of $O(\left|\Theta_{j}\right|^{3})$.
Dominated by the for-loop (lines 21-41), the time complexity of a
sub-auction (Algorithm~\ref{alg:CRM}) is $O(\left|\Theta_{j}\right|^{3})$.
Since $\sum_{j\in\{1,\ldots,G\}}\left|\Theta_{j}\right|=N$ and $\frac{N^{3}}{G^{2}}\leq\sum_{j\in\{1,\ldots,G\}}\left|\Theta_{j}\right|^{3}\leq N^{3}$,
the running time of the original RMA $M_{o}$ is bounded by polynomial
time $O(N^{3})$.
\end{IEEEproof}

\section{Deep Reinforcement Learning based Auction Mechanism (DRLA)\label{sec:DL_auction}}

Although the  RMA mechanism can guarantee the IC, IR and CE, its achieved
social welfare is still restricted. The reasons are that the randomization
may degrade the social welfare performance and the channel conflicts
among workers is not well represented and exploited. Resolving these
issues is very challenging. In this section, we attempt to utilize
the powerful artificial intelligence (AI) to establish an automated
mechanism for improving the social welfare while ensuring the IC and
IR. Specifically, we first use the graph neural network (GNN)~\cite{Scarselli2009}
to exploit the conflict relationships and generate effective embeddings.
Based on the embeddings, we propose a deep reinforcement learning
(DRL) framework to design truthful auction mechanisms in order to
improve the social welfare.  
\begin{figure}[tbh]
\begin{centering}
\includegraphics[width=0.95\columnwidth]{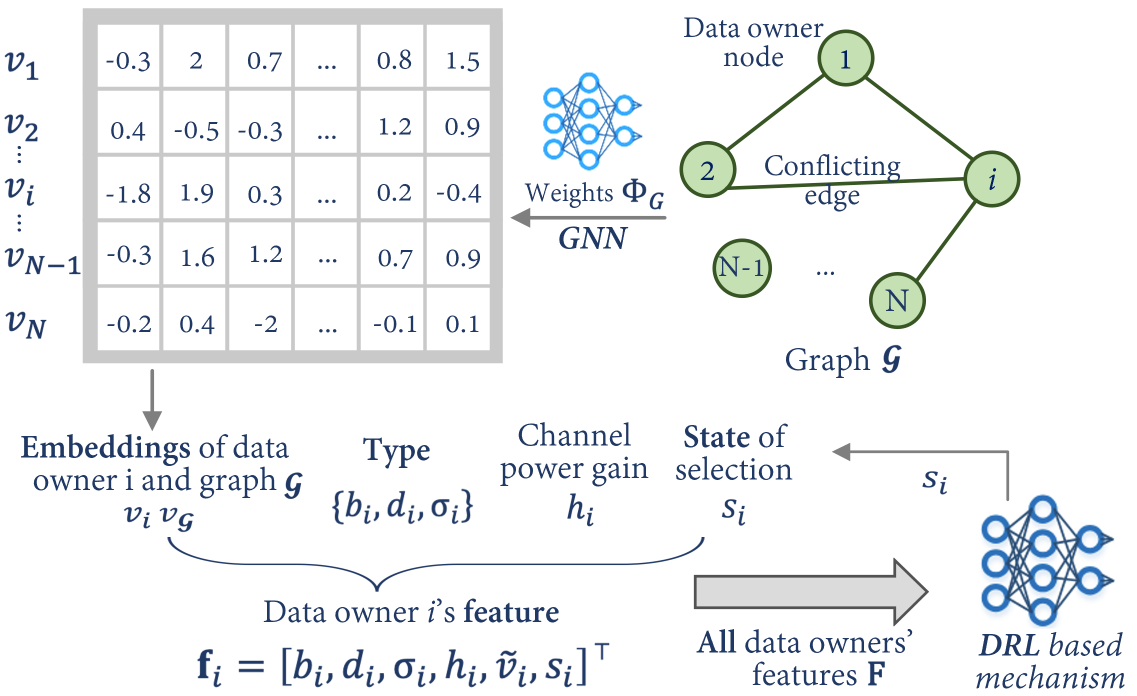}
\par\end{centering}
\caption{Feature engineering based on GNN.\label{fig:GNN}}
\end{figure}

\subsection{Feature engineering with embeddings of wireless spectrum conflict graph \label{subsec:Feature-engineering-conflict-graph}}

Although the bid $b_{i}$, data size $d_{i}$ and EMD $\sigma_{i}$
in data owner's type $\mathbf{t}_{i}$ and the channel information $h_{i}$
are already continuous variables, the information of requested wireless channels $\mathcal{C}_{i}$ is a discrete variable which restricts directly applying the DRL approach. Therefore, we construct a spectrum conflict graph $\mathcal{G}$~\cite{Zhou2015} to represent the channel conflicting relationship among the data owners. We here denote the data owner $i$'s \emph{original feature} by $\mathbf{f}^{\mathrm{o}}_{i}$, i.e.,  $\mathbf{f}^{\mathrm{o}}_{i}=[b_{i}, d_{i}, \sigma_{i}, h_{i}, \mathcal{C}_{i}]^{\top}$. $\top$ is the transpose operator and the square brackets $[\cdot]$ is the operators of incorporating the inside elements to a vector. 

As illustrated in Fig.~\ref{fig:GNN}, each node in the graph $\mathcal{G}$ is a data owner and each undirected edge represents the conflicting relationship between two connected data owners. Due to the differences in some aspects, such as hardware or wireless channel occupancy, each data owner may have different demands for wireless channels. Taking an example with $3$ data owners, the data owners $1,2$ and $3$ respectively request channels $\mathcal{C}{}_{1}=\{1,4,6\}$, $\mathcal{C}{}_{2}=\{2,5,6\}$
and $\mathcal{C}{}_{3}=\{3,7\}$. Since the data owners $1$ and $2$
are single-minded and both of them request the channel $6$, they
are conflicting in wireless channels and there should be an edge between data owners $1$ and $2$. The data owner $3$ has no channel conflicting with any other worker's requested channels, so there is no edge connected to data owner $3$.

To map the discrete channel information to continuous embeddings, we specifically apply a multi-layer Graph Convolutional Network (GCN)~\cite{Khalil2017} in which the $l+1$th layer output
$\mathbf{H}^{(l+1)}$ is calculated by 
\begin{equation}
\mathbf{H}^{(l+1)}=\mathrm{ReLU}(\mathbf{B}^{-\frac{1}{2}}\hat{\mathbf{A}}\mathbf{B}^{-\frac{1}{2}}\mathbf{H}^{(l)}\Phi_{G}^{(l)}),\label{eq:gvn_layer_operator}
\end{equation}
where $\mathbf{H}^{0}=\mathbf{1}^{N\times\varpi_{\mathrm{\mathcal{G}}}}$
is an all-ones matrix and $\hat{\mathbf{A}}=\mathbf{A}+\mathbf{I}$
denotes the adjacency matrix $\mathbf{A}\in\mathbb{N}^{N\times N}$
with self-connections. $\mathbf{I}\in\mathbb{N}^{N\times N}$ is an
identity matrix and $\mathbf{B}=\sum_{j=0}\hat{\mathbf{A}}_{ij}$ is
the diagonal degree matrix of $\hat{\mathbf{A}}_{ij}$. $\mathbf{\Phi}_{G}$ is the trainable parameter set of the GCN where $\Phi_{G}^{(l)}\in\mathbb{R}^{\varpi_{\mathrm{\mathcal{G}}}\times\varpi_{\mathrm{\mathcal{G}}}}$
is the trainable weight matrix of the $l$th layer. We use the rectified linear units $\mathrm{ReLU(\cdot)=}\max(0,\cdot)$~\cite{Nair2010} as an activation function. Then, the embedding $v_{i}\in\mathbb{R}^{\varpi_{\mathrm{\mathcal{G}}}\times1}$
of each data owner (node) $i$ generated by the GCN can be obtained
from the output of the last layer $\mathbf{H}^{(\hat{l}+1)}=[v_{1},v_{2},\ldots,v_{N}]^{\top}$, where $\hat{l}$ is the total number of layers. Based on the node embeddings, the embedding of the graph $\mathcal{G}$ is encoded by $v_{\mathcal{G}}=\sum_{i}v_{i}\in\mathbb{R}^{\varpi_{\mathrm{\mathcal{G}}}\times1}$.
We concatenate the node embedding and the graph embedding as the data owner $i$'s final embedding $\tilde{v}_{i}=[v_{i},v_{\mathcal{G}}]^{\top}\in\mathbb{R}^{2\varpi_{\mathrm{\mathcal{G}}}\times1}$.
Apart from the data owner's own type information, the state $s_{i}$
which is output by the DRL algorithm and indicates whether the worker $i$ wins the DRL based auction should be included in the input feature. If the data owner $i$ is selected to be included in the set of  workers, we set $s_{i}=1$; otherwise, $s_{i}=0$. Finally, the data owner $i$ is represented by its feature $\mathbf{f}_{i}=[b_{i},d_{i},\sigma_{i},C_{i},h_{i},\tilde{v}_{i},s_{i}]^{\top}\in\mathbb{R}^{(2\varpi_{\mathrm{\mathcal{G}}}+5)\times1}$
which incorporates its type, embeddings, state and channel information.
In addition, we use $\mathbf{F}=[\mathbf{f}_{1},\ldots,\mathbf{f}_{N}]^{\top}\in\mathbb{R}^{N\times(2\varpi_{\mathrm{\mathcal{G}}}+5)}$
 and $\mathbf{F}^{\mathrm{o}}=[\mathbf{f}^{\mathrm{o}}_{1},\ldots,\mathbf{f}^{\mathrm{o}}_{N}]$to respectively represent all data owners' features and original features. 

\subsection{Automated mechanism under deep Q-learning framework}

\begin{figure}[tbh]
\begin{centering}
\includegraphics[width=0.95\columnwidth]{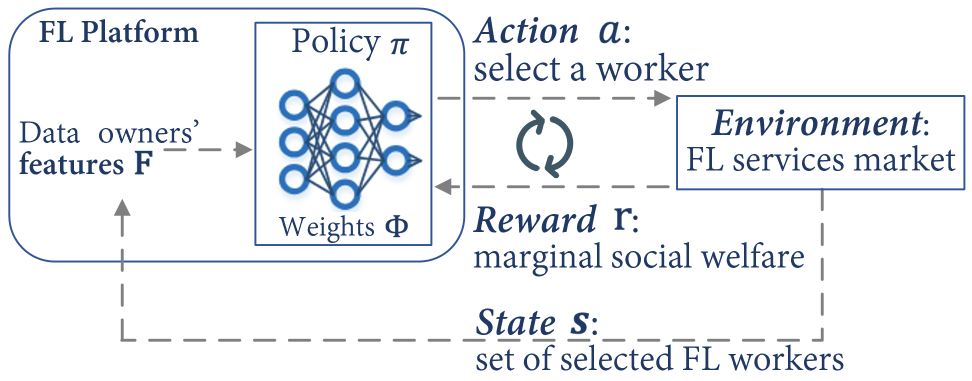}
\par\end{centering}
\caption{DRL based framework auction mechanism.\label{fig:DRL-auction}}
\end{figure}
Generally, we adopt the deep Q-learning~\cite{Khalil2017,Mnih2013}
framework to design an auction mechanism that possesses the properties of IC and IR and solves the NP-hard social welfare maximization problem.
Similar to the RMA, the DRL based auction mechanism applies the greedy
scheme which finds the  workers step by step. At the step $m$ (starting
from $1$), it selects a worker that has no channel conflict with
the candidate set $\mathcal{V}^{m}$ and maximizes an \emph{evaluation
function $Q$} on $\mathcal{V}^{m}$. After $\hat{m}$ steps reaching
the \emph{termination condition}, the worker selection process ends
and the final  worker set is $\hat{\mathcal{V}}=\mathcal{V}^{\hat{m}}$.
At step $1$, the initial candidate set is $\mathcal{V}^{1}=\emptyset$.

The core assumption for the DRLA mechanism is that the data owners' original features $\mathbf{F}^{\mathrm{o}}$ follow a distribution $\mathbb{D}$. When the service provider, i.e., the auctioneer, trains the DRLA network, it can obtain the data owners' original features from a historical real-world dataset or a priori known distribution~\cite{Khalil2017}. 

As illustrated in Fig.~\ref{fig:DRL-auction}, the proposed DRL framework
is composed of the state, action, reward, policy and environment parts
defined as follows:
\begin{itemize}
\item \emph{States}: the state $\mathbf{s}^{m}=\{s_{1}^{m},\ldots,s_{i}^{m},\ldots,s_{N}^{m}\}$
consists of the aforementioned workers' states at step $m$, indicating
whether they have joined the candidate set $\mathcal{V}^{m}$, i.e.,
$s_{i}^{m}=1$ if $i\in\mathcal{V}^{m}$ and $0$ otherwise. We use
function $V$ to express this transformation, $s^{m}=V(\mathcal{V}^{m})$.
\item \emph{Actions}: the action $a^{m}$ is a data owner, i.e., a node
in the graph $\mathcal{G}$, which is picked by the FL platform at
step $m$ and not in the candidate set $\mathcal{V}^{m}$. 
\item \emph{State transition}: the state transition from the current state
$\mathbf{s}^{m}$ to the next state $\mathbf{s}^{m+1}$ is determined
by the $m$th action which means setting $s_{a^{m}}=1$ and putting
the data owner (node) $a^{m}$ into the set $\mathcal{V}^{m}$, i.e.,
$\mathcal{V}^{m+1}=\mathcal{V}^{m}\cup\{a^{m}\}$.
\item \emph{Rewards}: the reward function $r^{m}(\mathbf{s}^{m},a^{m})$
at state $\mathbf{s}^{m}$ is the increased social welfare contributed
by the action $a^{m}$, which is defined as 
\begin{equation}
r^{m}(\mathbf{s}^{m},a^{m})=S(\mathcal{V}^{m}\cup\{a^{m}\})-S(\mathcal{V}^{m})\label{eq:marginal-sw}
\end{equation}
where $S(\cdot)$ is the social welfare function in (\ref{eq:sw_maixmization}).
Then, the cumulative reward $R=\sum_{m=1}^{\hat{m}}r^{m}(\mathbf{s}^{m},a^{m})$
is equal to our optimization target, i.e., the final achieved social
welfare $S(\mathcal{V}^{\hat{m}})$. 
\item \emph{Policy}: different from the traditional Q-learning~\cite{Sutton2018},
which uses a Q-table, the adopted DRL trains a deep neural network
(DNN) $Q(\mathbf{s}^{m},a^{m}|\mathbf{\Phi}_{Q})$ to evaluate the
quality of action $a^{m}$ under state $\mathbf{s}^{m}$. That is,
the input of DNN is any state and action pair and the output is the
corresponding quality value. $\mathbf{\Phi}_{Q}$ denotes the trainable parameter set of the DNN. Thus, $\mathbf{\Phi}=[\mathbf{\Phi}_{G}, \mathbf{\Phi}_{Q}]$ represents all trainable parameters of the proposed \emph{DRLA networks}. Based on the evaluation function $Q$, we
use the greedy policy $\pi(a^{m}|\mathbf{s}^{m})=\arg\max_{a^{m}\in\mathcal{N}\setminus(\mathcal{V}^{m}\cup\mathcal{L}(\mathcal{V}^{m}))}Q(\mathbf{s}^{m},a^{m})$
to choose the action $a^{m}$ under state $\mathbf{s}^{m}$. In the
specific algorithm, we apply $\varsigma$-greedy policy. That is,
the DRL based auction at step $m$ randomly chooses a worker from
$\mathcal{V}^{m}$ with probability $\varsigma$, or implement the
policy $\pi(a^{m}|\mathbf{s}^{m})$ with probability $1-\varsigma$.
\end{itemize}

With the Q function in the classical deep Q-learning~\cite{Khalil2017, mnih2015human}
and the prepared data owners' features $\mathbf{F}$, the DNN based
evaluation function $Q(\mathbf{s}^{m},a^{m}|\mathbf{\Phi}_{Q})$ is
designed as
\begin{align}
Q(\mathbf{s}^{m},a^{m}|\mathbf{\Phi}) &= Q(\mathbf{s}^{m},a^{m}|\mathbf{\Phi}_{G}, \mathbf{\Phi}_{Q}) \\& =\mathrm{ReLU}([\tilde{\mathbf{v}},\mathbf{s},\mathbf{C},\mathbf{\mathbf{h}}]^{\top}\Phi_{Q}^{1})\Phi_{Q}^{2}\nonumber \\
 & \quad\quad\quad\quad\quad\quad\quad-\mathbf{b}^{\top}\mathrm{e}^{\Phi_{Q}^{3}}+g(\mathbf{d},\boldsymbol{\sigma})\mathrm{e}^{\Phi_{Q}^{4}})\label{eq:Q-function}
\end{align}
where $\tilde{\mathbf{v}}=[\tilde{v}_{1},\ldots,\tilde{v}_{N}]^{\top}$,
$\mathbf{s}=[s_{1},\ldots,s_{N}]^{\top}$, $\mathbf{C}=[C_{1},\ldots,C_{N}]^{\top}$,
$\mathbf{\mathbf{h}}=[h_{1},\ldots,h_{N}]^{\top}$, $\mathbf{b}=[b_{1},\ldots,b_{N}]^{\top}$,
$\mathbf{d}=[d_{1},\ldots,d_{N}]^{\top}$ and $\boldsymbol{\sigma}=[\sigma_{1},\ldots,\sigma_{N}]^{\top}$.
Besides the parameters $\Phi_{Q}^{1}\in\mathbb{R}^{(2\varpi_{\mathrm{\mathcal{G}}}+3)\times(2\varpi_{\mathrm{\mathcal{G}}}+3)}$,
$\Phi_{Q}^{2},\Phi_{Q}^{3}\in\mathbb{R}^{(2\varpi_{\mathrm{\mathcal{G}}}+3)\times1}$
and $\Phi_{Q}^{4}\in\mathbb{R}$, the evaluation function integrates
a monotonic neural network function $g(\mathbf{d},\boldsymbol{\sigma})\in\mathbb{R}$
described by
\begin{align}
g(\mathbf{d},\boldsymbol{\sigma}) & =\max_{j\in\{1,\ldots,J\}}\min_{k\in\{1,\ldots,K\}}\{\mathrm{ReLU}(\mathrm{ReLU}(\nonumber \\
 & \quad\quad\quad[\mathbf{d},-\boldsymbol{\sigma}]^{\top}\mathrm{e}^{\Phi_{Q}^{5_{jk1}}}+\Phi_{Q}^{6_{jk1}})\mathrm{e}^{\mathbf{\Phi}_{Q}^{5_{jk2}}}+\Phi_{Q}^{6_{jk2}})\}\label{eq:monotonic-function}
\end{align}
where $J$ and $K$ are positive integral hyper-parameters that adjust
the approximate accuracy and the complexity of the monotonic network
$g$, $\Phi_{Q}^{5_{jk1}}\in\mathbb{R}^{2\times K}$, $\Phi_{Q}^{6_{jk1}}\in\mathbb{R}^{N\times K}$
are the parameters in the first layer, and $\Phi_{Q}^{5_{jk2}}\in\mathbb{R}^{K\times1}$,
$\Phi_{Q}^{6_{jk2}}\in\mathbb{R}^{N\times1}$ are the parameters in
the second hidden layer. The exponential operations in (\ref{eq:Q-function})
and (\ref{eq:monotonic-function}), i.e., $\mathrm{e}^{\Phi_{Q}^{3}}$,
$\mathrm{e}^{\Phi_{Q}^{4}}$, $\mathrm{e}^{\Phi_{Q}^{4_{jk1}}}$ and
$\mathrm{e}^{\mathbf{\Phi}_{Q}^{4_{jk2}}}$, guarantee that the coefficients of input features $-\mathbf{b}$, $[\mathbf{d},-\boldsymbol{\sigma}]$
and $g$ are positive. According to the characterizations of the monotonic
network in~\cite{Sill1998,You2017}, $g(\mathbf{d},\boldsymbol{\sigma})$
and $Q(\mathbf{s}^{m},a^{m}|\mathbf{\Phi})$ are monotonically
decreasing with the data owner's EMD value and monotonically increasing with the data size. $Q(\mathbf{s}^{m},a^{m}|\mathbf{\Phi})$ is clearly monotonically decreasing with bid. The DNN $Q$ can be seen as a scoring function~\cite{Bichler2005,Asker2008} that calculates the data owner $i$'s score $Q_{i}$. This is convenient for the FL platform to sort and select the  data owners. Actually, $g(\mathbf{d},\boldsymbol{\sigma})$ maps the data owner $i$'s data size $d_{i}$ and EMD $\sigma_{i}$ to a new metric value $g_{i}$. Analogous to the RMA, the setting in the DRLA mechanism is transformed to that each data owner wants to sell $g_{i}$ units of data at price $b_{i}$.
$g_{i}$ can be regarded as the data owner $i$'s \emph{normalized
data size}. Thus, the truthfulness and individual rationality of the
DRLA mechanism are also guaranteed by Theorem~\ref{thm:multi_unit_charac} where we accordingly replace $d_{i}$ with $g_{i}$. With the FL platform's greedy policy, the trained deep Q-learning model with fixed parameters always chooses the maximum $Q_{i}$ at each step and thus possesses the monotonicity in worker selection. The payment $p_{i}$ is set to the  worker $i$'s critical bid, i.e., the maximum bid $\bar{b}_{i}$ that keeps $Q_{i}$ as a winning score. Similar to the RMA (lines 21-41), the payment determination is applying the deep Q-learning model on the set of data owners except the worker $i$ to find the critical  data owner $\bar{i}$ which is the last data owner selected or the first one having channel conflicts with worker $i$. Due to the monotonicity of the bid in $Q$ function, the FL platform can efficiently calculate the bid $\bar{b}_{i}$ such that worker's new score $\bar{Q}_{i}$ is equal to the critical data owner $\bar{i}$'s
score $Q_{\bar{i}}$.
\begin{algorithm}[tbh]
{\scriptsize{}\footnotesize
\begin{algorithmic}[1] 
\Require{Distribution $\mathbb{D}$ or real dataset}
\Ensure{The set of FL workers $\mathcal{W}$.} 
\State{Initialize parameters $\mathbf{\Phi}$, $\mathbf{\Phi}^-$, experience replay memory $\mathcal{M}$}
\Begin 
	\For{episode $k=1$ to $L_1$}
		\State{Draw $\mathbf{F}^{\mathrm{o}}$ from distribution $\mathbb{D}$ or real dataset}
		\State{Initialize the candidate set to be empty $\mathcal{V}^{m}=\emptyset$}
		\For{step $m=1$ to $L_2$}
			\State{$\mathcal{\widehat{V}}^{m} \gets \mathcal{V}^{m}\cup\mathcal{L}(\mathcal{V}^{m})$}
			\State{$a^{m}=\begin{cases}\mathrm{randomly\,choose}\,a^{m}\in\mathcal{N}\setminus\mathcal{\widehat{V}}^{m}, & w.p.\,\varsigma\\\arg\max_{a^{m}\in\mathcal{N}\setminus\mathcal{\widehat{V}}^{m}}Q(\mathbf{s}^{m},a^{m}), &w.p.\,1-\varsigma\end{cases}$}
			\State{$\mathcal{V}^{m} \gets \mathcal{V}^{m}\cup \{a^m\}$}
			\State{$\mathbf{s}^{m+1}\gets V(\mathcal{V}^{m})$}
			\State{Execute action $a^m$ to obtain reward $r^m$}
			\If{$r^{m}<0$ \textbf{or} $\mathcal{N}\setminus\mathcal{\widehat{V}}^{m}=\emptyset$}
				\Break
			\EndIf
			\If{$m\geq\mu_{\mathrm{u}}$}
				\State{Store $\{s^{m-\mu_{\mathrm{u}}},a^{m-\mu_{\mathrm{u}}}, r^{m-\mu_{\mathrm{u}}}, s^{m}, \mathbf{F}\}$ in $\mathcal{M}$}
				\State{Sample minibatch $\mathcal{B}$ from memory $\mathcal{M}$}
				\State{Update $\mathbf{\Phi}$ by SGD over (\ref{eq:square-loss})}
			\EndIf
		\EndFor
		\If{$k$ $\mathrm{mod}$ $\mu_{\mathrm{r}}=0$}
			\State{$\mathbf{\Phi}^{-} \gets \mathbf{\Phi}$}
		\EndIf
	\EndFor
\End
\end{algorithmic}\caption{DRLA training algorithm\label{alg:DRL-auction}}
}
\end{algorithm}

The training process of our proposed DRL based auction mechanism is
presented in Algorithm~\ref{alg:DRL-auction}. At the beginning
of each episode, the platform first samples a set of worker's original features $\mathbf{F}^{\mathrm{o}}$ from a known distribution $\mathbb{D}$ or a real-world dataset. Training the DNN, i.e., the evaluation function, can help the FL platform to establish the optimal policy that finds the best action at the current state. The standard Q-learning updates the $Q$ function parameters based on the immediate reward $r^{m}$ at the
$m$th step of an episode. However, the standard update method is
myopic since our objective is to optimize the total social welfare,
i.e., the accumulated reward $R$. Thus, our deep Q-learning training
shifts to use the additive reward from the past $\mu_{\mathrm{u}}$
steps, i.e., $R_{\mu_{\mathrm{u}}}^{m}=\sum_{m-\mu_{\mathrm{u}}}^{m}r^{m}$.
To improve the training stability, an experience replay memory $\mathcal{M}$
is created for storing the experiences, e.g., $\left\{ s^{m-\mu_{\mathrm{u}}},a^{m-\mu_{\mathrm{u}}},R_{\mu_{\mathrm{u}}}^{m},s^{m},\mathbf{F}\right\} $.
Moreover, a single DNN may also lead to the overestimation~\cite{VanHasselt2016}
since the FL platform's action is selected and evaluated by the same
$Q$ function. To address this issue, we apply the double deep Q learning
(DDQL)~\cite{VanHasselt2016}. Specifically, we have two DRLA networks, including the original evaluation DRLA network with parameters $\mathbf{\Phi}$ and
an additional target DRLA network with parameters $\mathbf{\Phi}^{-}$. The parameters $\mathbf{\Phi}$ of the evaluation DNN can be updated
by using the gradient descent at each step after $\mu_{\mathrm{u}}$
steps in an episode to minimize the following square loss function:
\begin{equation}
\mathrm{loss}=(\hat{Q}-Q(\mathbf{s}^{m-\mu_{\mathrm{u}}},a^{m-\mu_{\mathrm{u}}}|\mathbf{\Phi}))^{2}. \label{eq:square-loss}
\end{equation}
The target value $\hat{Q}$ is defined as 
\begin{equation}
\hat{Q}=R_{\mu_{\mathrm{u}}}^{m}+\lambda^{\mu_{\mathrm{u}}}(Q(\mathbf{s}^{m},\arg\max_{a^{m}\in\mathcal{N}\setminus\mathcal{\widehat{V}}^{m}}Q(\mathbf{s}^{m},a^{m}|\mathbf{\Phi})|\mathbf{\Phi}^{-})),
\end{equation}
where $\mu_{\mathrm{u}}\leq m\leq\hat{m}$ and $\lambda$ is a discount
factor. The target DNN resets its parameters $\mathbf{\Phi}^{-}=\mathbf{\Phi}$
at every $\mu_{\mathrm{r}}$ episodes. The termination condition for
the training in an episode is that the immediate reward $r^{m}$ becomes
negative, i.e., $r^{m}<0$, or there is no worker to select, i.e.,
$\mathcal{N}\setminus\mathcal{\widehat{V}}^{m}=\emptyset$. For higher
robustness of convergence, we use stochastic gradient descent (SGD)
to train the evaluation DNN over a minibatch $\mathcal{B}$ of $\mu_{\mathrm{B}}$
experiences randomly drawn from memory $\mathcal{M}$. The proposed DRLA mechanism actually adopts the classical DRL framework proposed in~\cite{Mnih2013} which has the stable convergence in training large neural networks using the reinforcement learning signal and the SGD method. In Section~\ref{sec:Experimental-and-simulation}, we present the experimental result on the convergence in training the DRLA network.

\section{Experimental Results and Discussions\label{sec:Experimental-and-simulation} }

In this section, we first conduct a federated learning experiment
based on real-world data to verify the proposed data utility function.
From the simulation results, we then examine the performance of the
proposed RMA and DRLA mechanisms.

\subsection{Verification for Data Utility Function\label{subsec:Veri-for-Data-util}}

\begin{figure}[tbh]
\begin{centering}
\includegraphics[width=0.85\columnwidth]{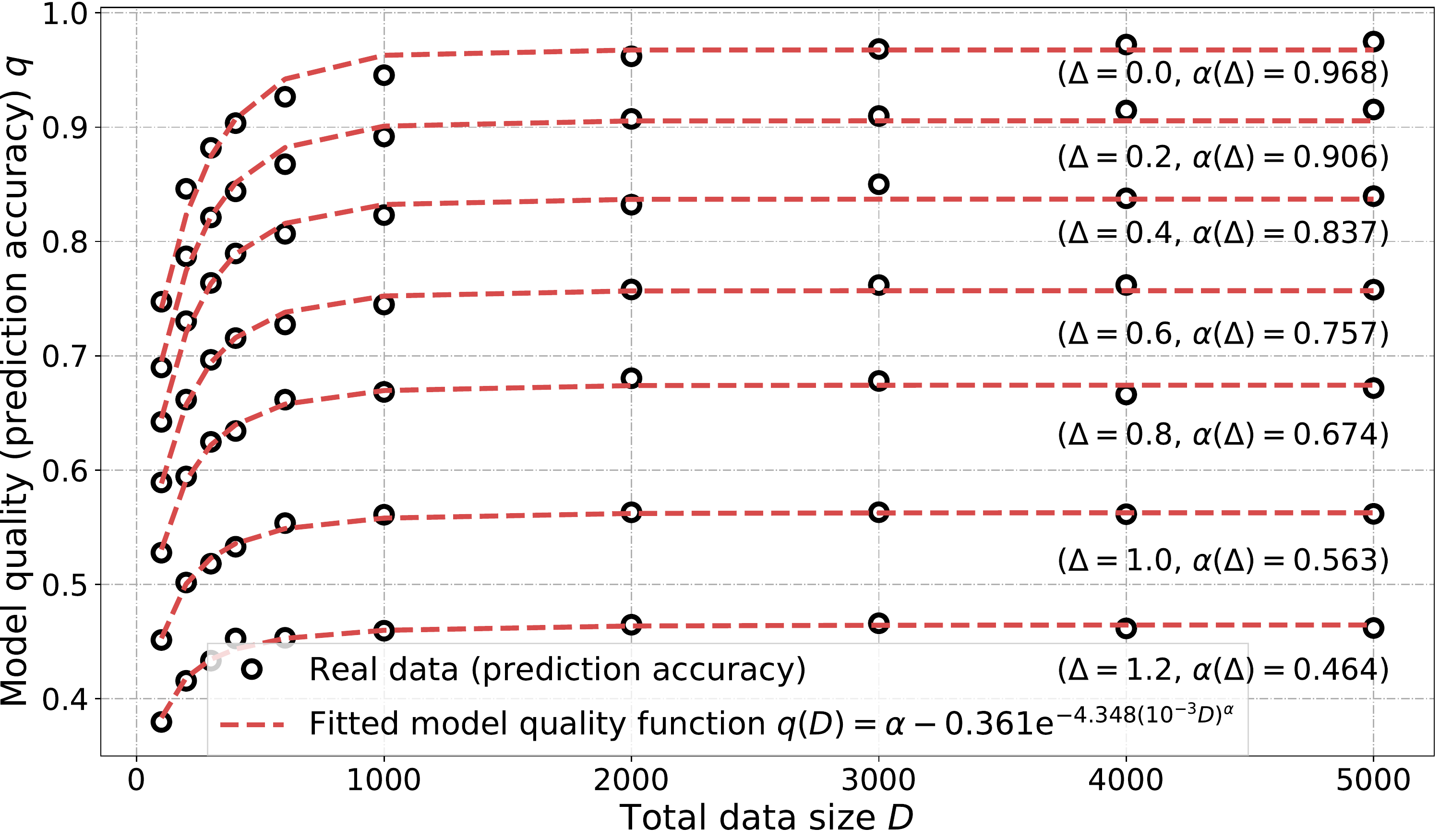}
\par\end{centering}
\bigskip{}

\begin{centering}
\includegraphics[width=0.85\columnwidth]{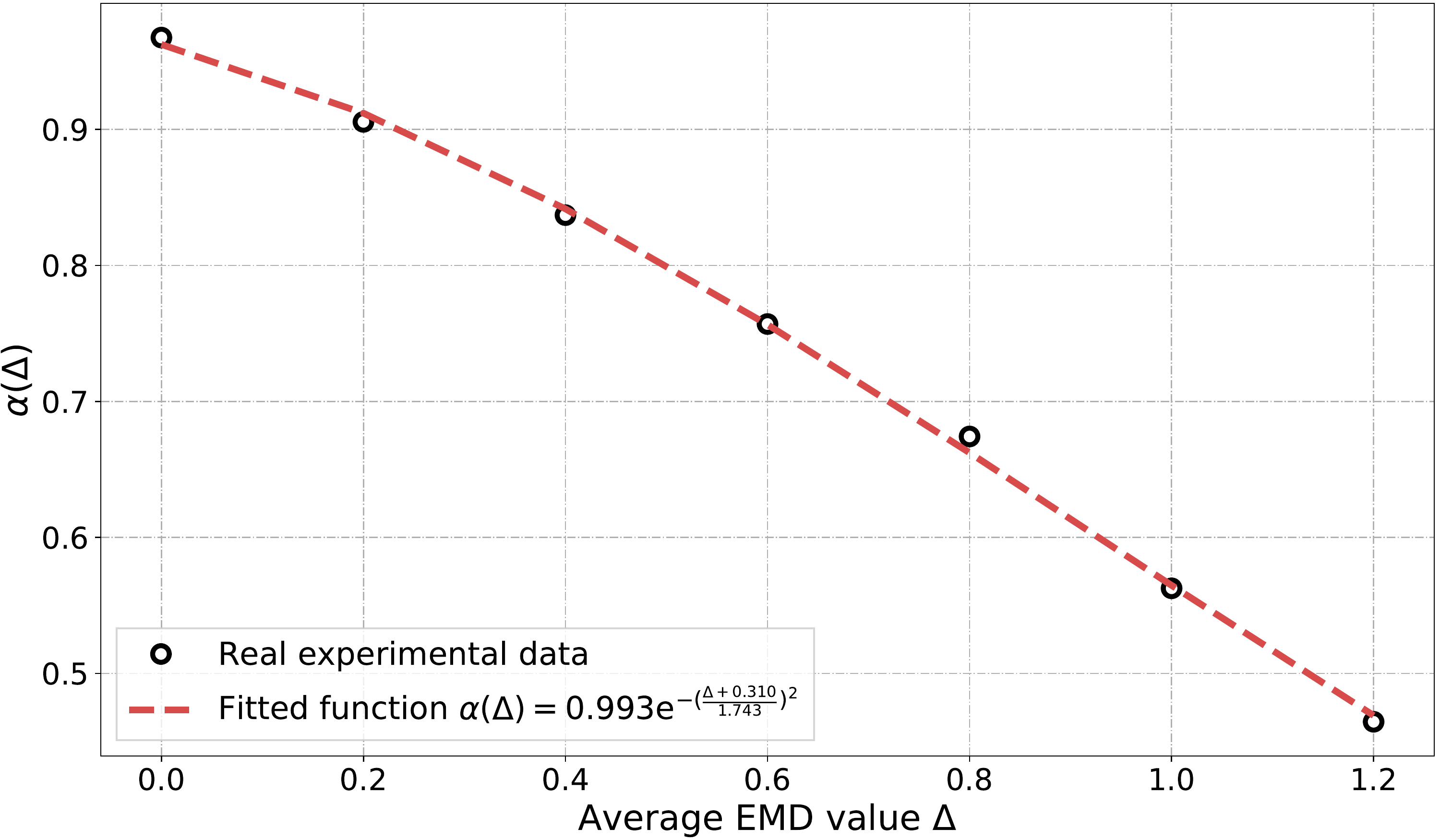}
\par\end{centering}
\caption{Estimation of the data utility function in (\ref{eq:quality_func}).\label{fig:estimation-of-data-utility}}

\end{figure}
To verify the data utility function defined in (\ref{eq:quality_func}),
we use the convolutional neural network (CNN) model on the classical
MNIST dataset\footnote{http://yann.lecun.com/exdb/mnist/} to develop
a federated handwritten digit recognition service. For simplicity
in our experiments, we first consider $2$ workers following the FedAvg
algorithm (Algorithm~\ref{alg:AveragingAlg}) and cooperating to
train the CNN model with two convolutional and two fully-connected
layers. The federated learning rate is $\eta=0.01$ and the number
of global epochs and local epochs are fixed at $\delta_{g}=10$ and
$\delta_{\mathrm{l}}=5$. The MNIST dataset contains $60,000$ training
samples and $10,000$ testing samples for $10$ digit labels from
$0$ to $9$. It is reasonable to assume that each label essentially
has an equal occurrence probability. So we set the actual distribution
for the whole population $\mathbb{P}$, i.e., the benchmark for measuring
the EMD value, as $\mathbb{P}(y=j)=0.1,\forall j\in\{0,\ldots,9\}$.
The worst EMD $\sigma_{\max}$ that the FL platform can accept is
set to be $1.2$, i.e., $\sigma_{\max}=1.2$. We vary the normalized
total data size $D$ and the average EMD value $\Delta$ by changing
each worker $i$' local data size and number of labels. Each presented
result is averaged over $100$ instances. The data utility here is
the prediction accuracy. Fig.~\ref{fig:estimation-of-data-utility}
demonstrates that the data utility function in (\ref{eq:quality_func})
well fits the real experiment results. Based on the experiment, we
set $\kappa_{1}=0.361$, $\kappa=4.348$, $\kappa_{3}=10^{-3}$, $\kappa_{4}=0.993$,
$\kappa_{5}=0.31$, $\kappa_{6}=1.743$, $\kappa_{7}=100$, $\delta_{g}=10$,
$\delta_{\mathrm{l}}=5$ and $M=0.5$ in the following simulations.

\subsection{Performance of RMA and DRLA mechanisms}

\begin{figure}[tbh]
	\begin{centering}
	\includegraphics[width=0.85\columnwidth]{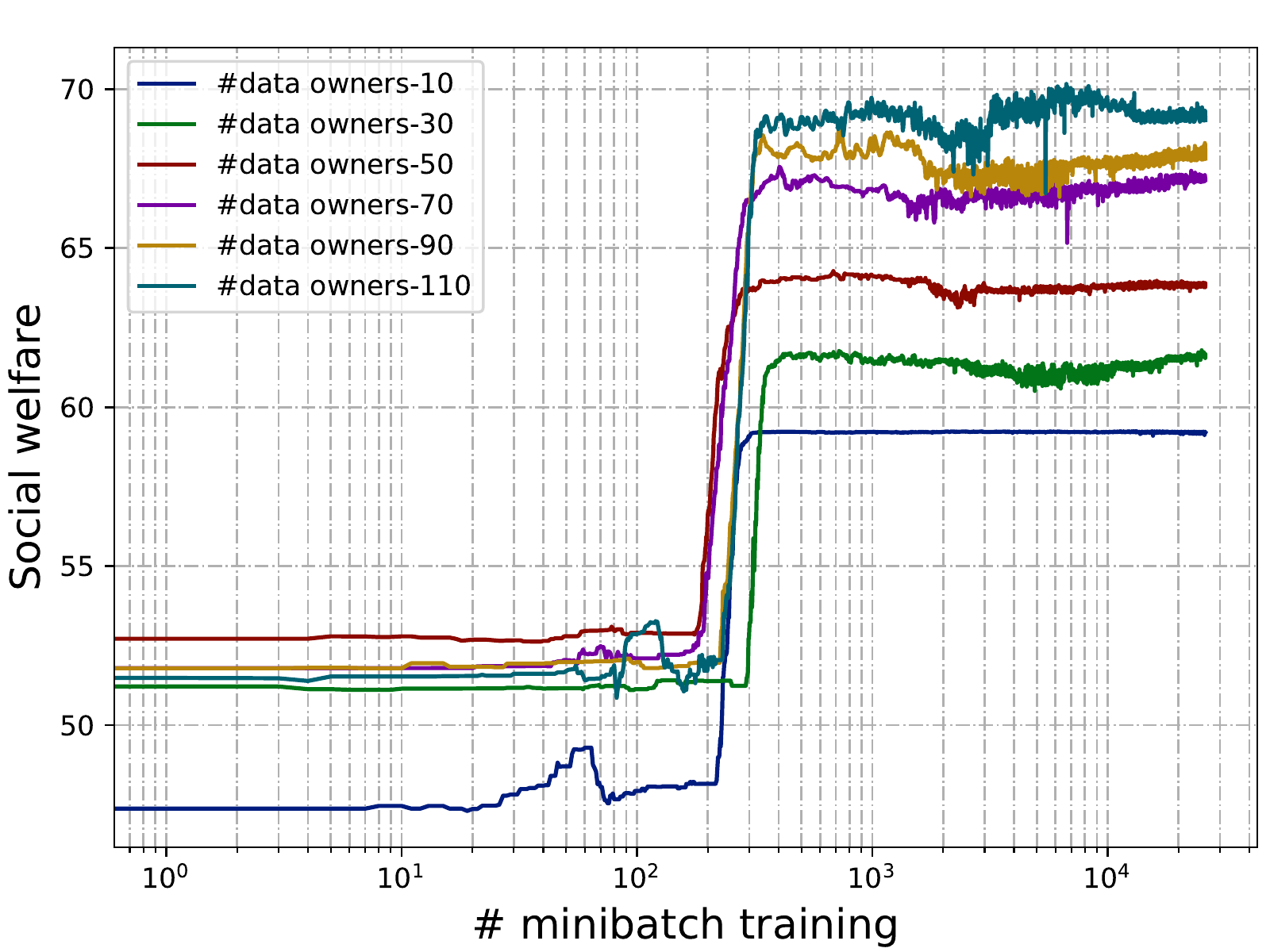}
	\par\end{centering}
	\caption{DRLA convergence measured by the held-out validation performance.\label{fig:convergence}}
\end{figure}

We conduct simulations to evaluate the performance of our proposed
strategyproof auction mechanisms, including the manually designed
RMA and the automated DRLA. Unless otherwise stated, the simulation
parameters are configured as follows. There are $N=50$ data owners
joining in the auction for participation in the federated learning
activity. We assume that the noise power spectral density level $\psi_{0}$
is $-130$ dBm/Hz, and the dynamic range of the channel power gain
$\tilde{h}^{2}$ is from $-90$ dB to $-100$ dB. Hereby, we uniformly
generate data owner $i$'s normalized channel power gain $h_{i}$
from $\left[10^{6},10^{7}\right]$, data size $d_{i}$ from $\left[0,10\right]$,
EMD value $\sigma_{i}$ from $\left[0,1.2\right]$, both unit data
collection cost $\gamma_{i}$ and unit data computational cost $\alpha_{i}$
from $\left[10^{-5},10^{-4}\right]$, and unit data transmission cost
from $\left[10^{-2},10^{-1}\right]$. Here, the worker's maximum data
size $d_{\max}=10$ and maximum EMD value $\sigma_{\max}=1.2$. The
platform's unit costs for computing and transmission are set as $\hat{\alpha}=5\times10^{-2}$
and $\hat{\beta}=5\times10^{-5}$. With respect to the wireless channels,
we fix the \emph{average number of channels per worker} $\bar{C}$
at $2$ and then the set of total available channels is $\mathcal{C}=\{1,\ldots,100\}$.
Each data owner' requested channel set $\mathcal{C}_{i}$ are uniformly
sampled from $\mathcal{C}$ and the corresponding number of channels
$C_{i}$ also follows uniform distribution in $\left[2,6\right]$.
We prepare $5,000$ samples for the DRLA model training, $100$ samples
for validation and $1,000$ samples for testing and evaluating the
performance of both the RMA and DRLA mechanisms. For RMA mechanism,
we set the number of groups as $G=10$. We implement the DRLA mechanism
integrated with a $2$-layer GCN and a monotonic network where $K=8$, $J=8$ and $\varpi_{\mathrm{\mathcal{G}}}=64$. We use the ADAM optimizer~\cite{kingma2014adam}
with a learning rate of $0.001$ and minibatch of $128$ and linearly anneal the exploration probability $\varsigma$ from $0.9$ to $0.05$ when training the DRLA model. As illustrated in Figure~\ref{fig:convergence}, we plot our proposed DRLA mechanism's convergence curves with respect to the held-out validation performance for different number of data owners. The performance of DRLA mechanism, i.e., the achieved social welfare or the accumulated reward, can quickly converge to a stable high value after training with a few hundreds of minibatches. All the following experimental results are the mean values based on the $1,000$ testing samples.


\begin{figure}[tbh]
\begin{centering}
\includegraphics[width=0.85\columnwidth]{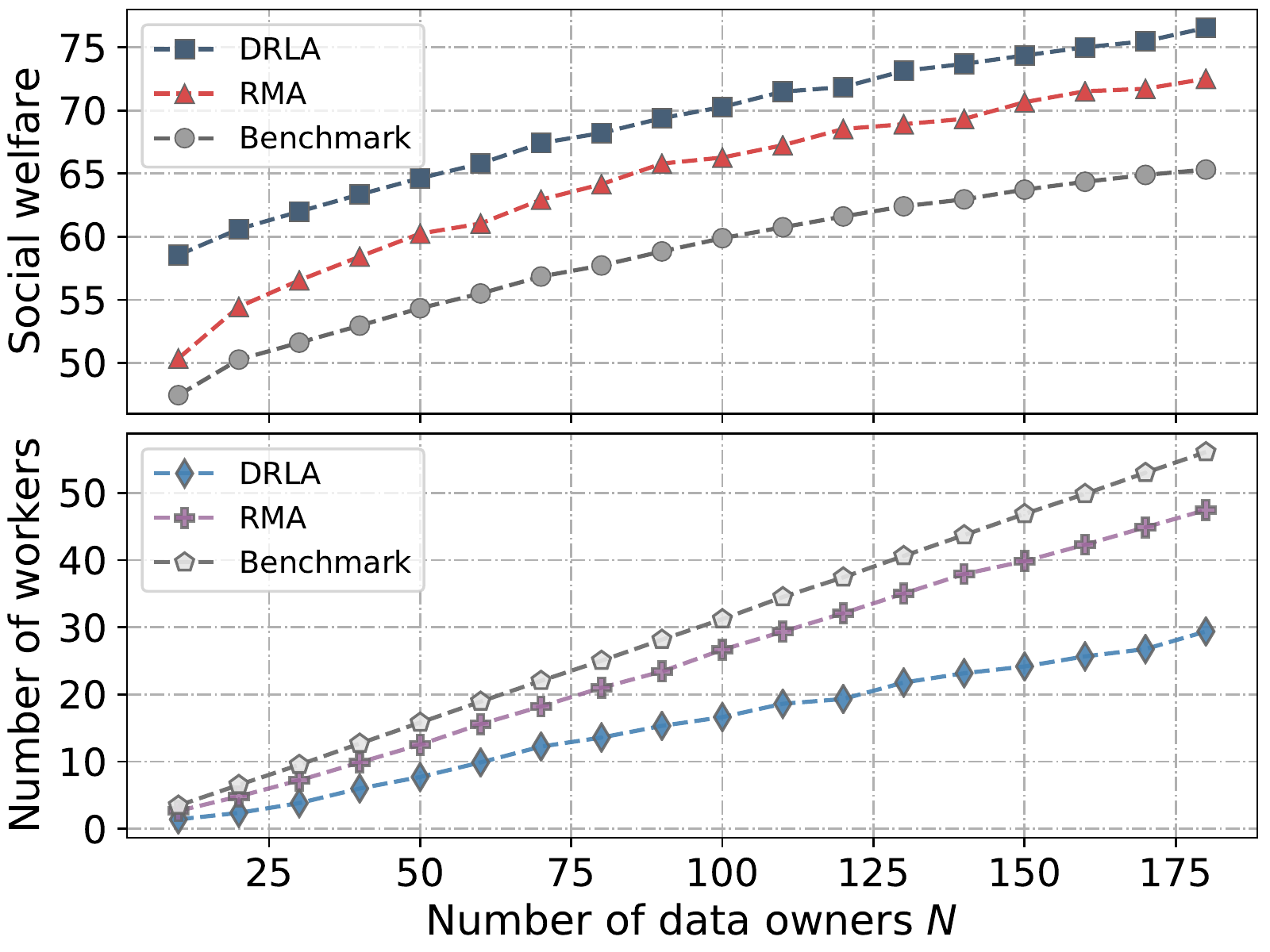}
\par\end{centering}
\caption{Impact of number of data owners $N$.\label{fig:Impact-of-N}}
\end{figure}

\begin{figure}[tbh]
\begin{centering}
\includegraphics[width=0.85\columnwidth]{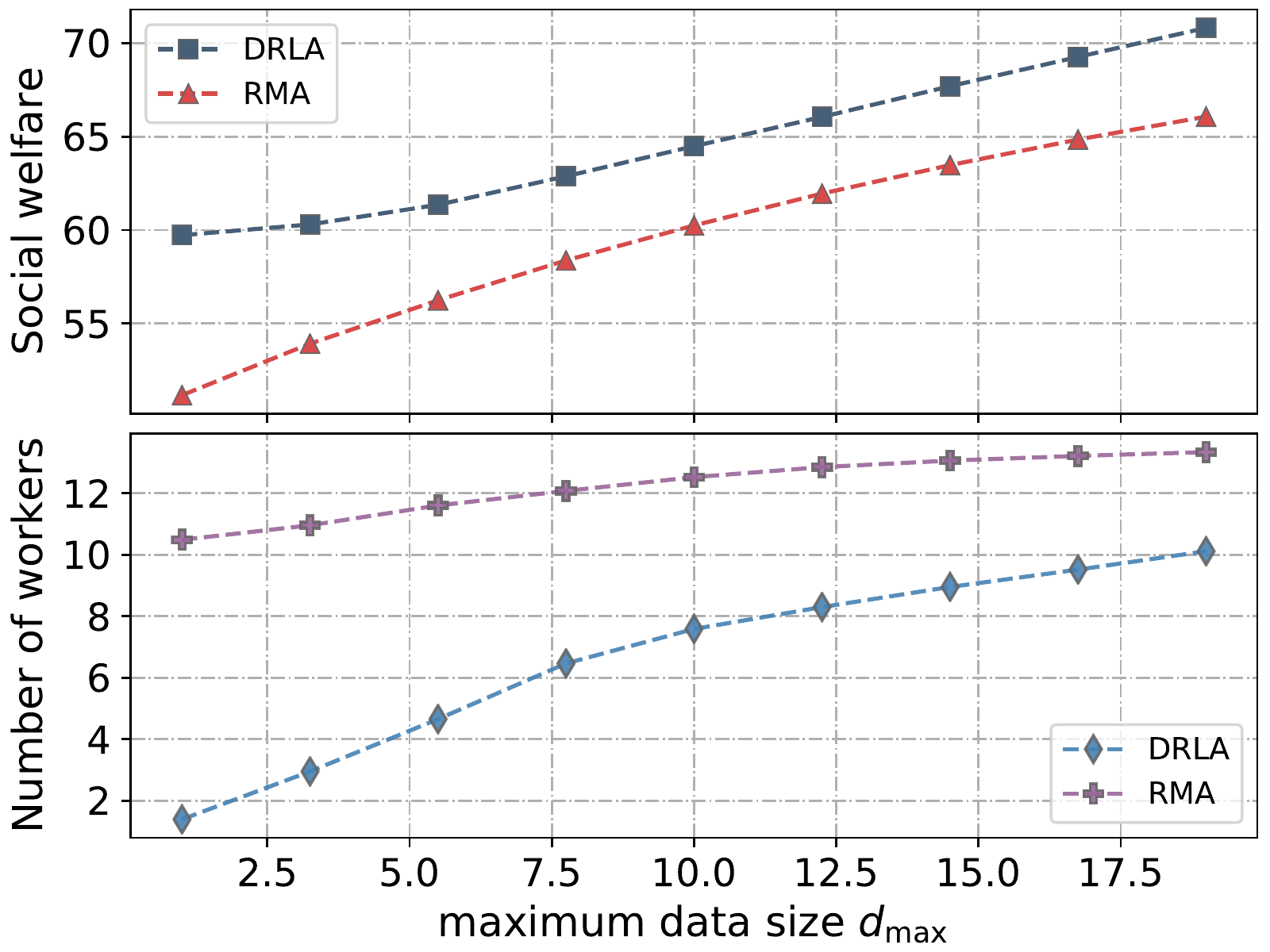}
\par\end{centering}
\caption{Impact of maximum data size $d_{\max}$.\label{fig:Impact_of_d_max}}
\end{figure}
\begin{figure}[tbh]
\begin{centering}
\includegraphics[width=0.85\columnwidth]{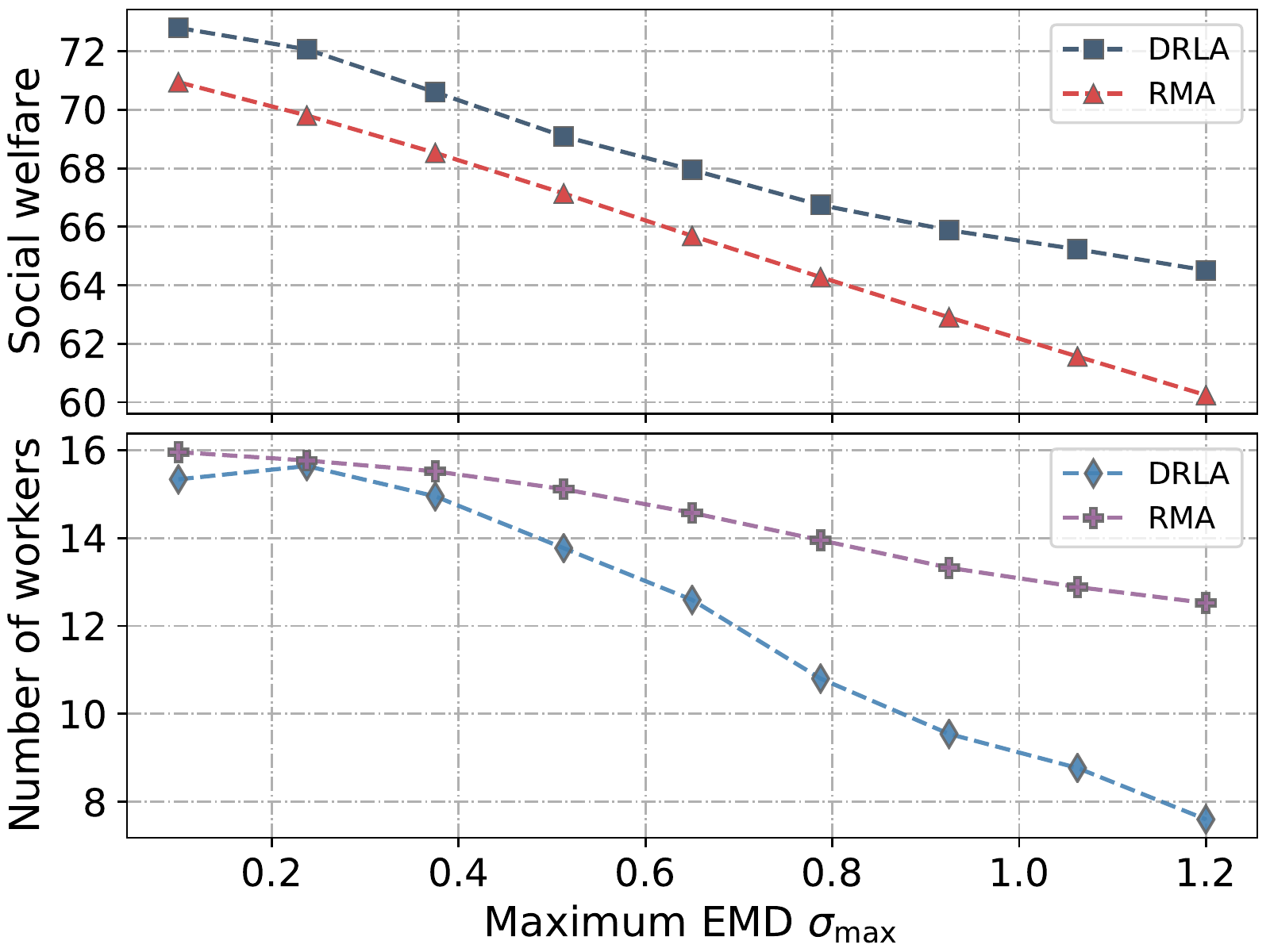}
\par\end{centering}
\caption{Impact of maximum EMD $\sigma_{\max}$.\label{fig:Impact_of_emd_max}}
\end{figure}

In addition to the social welfare metric, we are interested in the number of selected workers $W$ which reflects the fairness and the satisfaction rate of data owners. Figure~\ref{fig:Impact-of-N} demonstrates the impact of the number of data owners $N$ on the social welfare $S$ and the number of workers $W$. We observe that the social welfare and number of workers in the RMA and DRLA mechanisms both increase with growing data owners at a diminishing rate. The reason is two folds: First, the greedy algorithms only choose the worker that can improve the social welfare. Second, a larger base of interested workers will bring more competition in the auction and that more workers would also reduce the remaining workers' marginal social welfare density. Although the DRLA can achieve higher social welfare than that of RMA mechanism, the DRLA is less fair since it is more capable of sequentially finding out the data owner with larger marginal social welfare. As mentioned in Section~\ref{sec:Introduction}, there is few research work discussing the auction mechanism dedicated for wireless federated learning. Since the key challenging issue in this paper is about the channel conflict, we attempt to apply a well-known strategyproof spectrum auction mechanism proposed in~\cite{Zhou2008} as the benchmark. The benchmark auction mechanism decides the allocation only based on the bidders' bid prices while avoiding the channel conflicts among the bidders. As shown in the Figure~\ref{fig:Impact-of-N}, the social welfare achieved by the benchmark auction mechanism is lower than the proposed auction mechanisms, although it provides better fairness. However, this paper focuses on the social welfare optimization and thus the benchmark mechanism is not suitable to be directly applied in our wireless federated learning scenario. 

\begin{figure}[tbh]
\begin{centering}
\includegraphics[width=0.85\columnwidth]{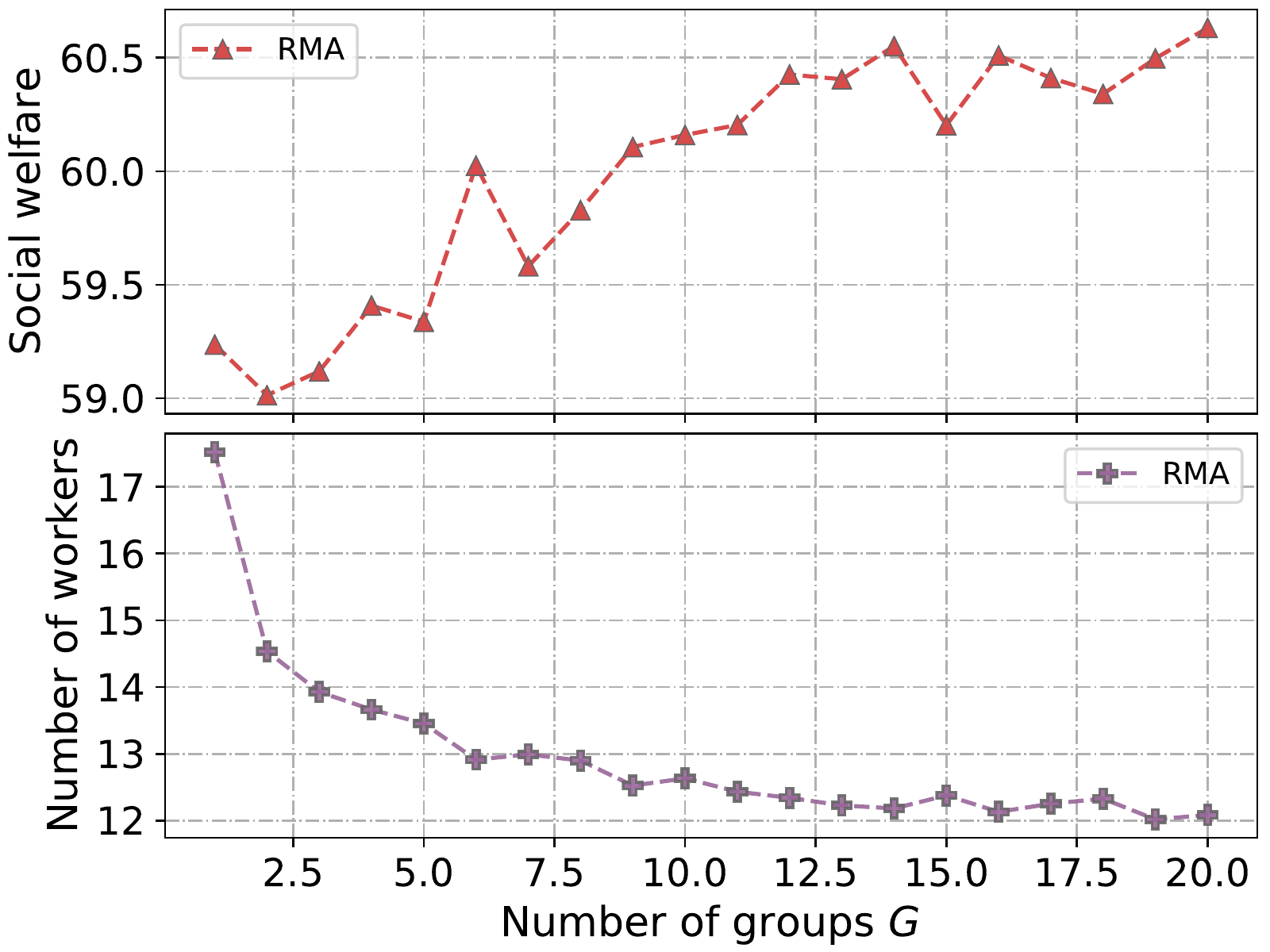}
\par\end{centering}
\caption{Impact of the number of groups $G$ on the RMA mechanism performance.\label{fig:CRM_G}}
\end{figure}


As illustrated in Figs.~\ref{fig:Impact_of_d_max} and~\ref{fig:Impact_of_emd_max},
we investigate the impact of the worker's maximum data size $d_{\max}$
and EMD value $\sigma_{\max}$ on the social welfare. Note that $d_{\max}$
and $\sigma_{\max}$ are adjustable parameters and preset by the FL
platform before the auction. It is clear that the social welfare increases
when the FL platform raises its requirement of data quality by larger
data size and lower EMD value. Certainly, the precondition is that
there are enough data owners that satisfy the requirement. It is interesting
to note that when $\sigma_{\max}$ is large, the DRLA can drop more
data owners with low data quality (high $\sigma_{i}$) to keep better
social welfare than that of the RMA. In Fig.~\ref{fig:CRM_G}, we
vary the number of groups $G$ to show its impact on the performance
of the RMA mechanism. With the $G$ growing, the achieved social welfare
is increasing while less workers are selected. More groups means the
virtual EMD difference among data owners is widening, so the RMA mechanism
can recognize more data owners with low original EMD value. When such
data owner with high data quality is found, there is less need to
choose other data owners with poor data quality. 

\section{Conclusion\label{sec:Conclusion}}

In this paper, we have proposed an auction based market model for
trading federated learning services in the wireless environment. We
have designed a reverse multi-dimensional auction (RMA) mechanism
for maximizing the social welfare of the federated learning services
market. The RMA mechanism not only considers workers' bid prices for
providing training services but also takes each worker's own multiple
attributes, including the data size, EMD, and wireless channel demand,
into account. To well evaluate each workers' value, we have introduced
a data quality function verified by real world experiments to characterize
the relationship between the accuracy performance and the size and
average EMD of all local data. To further improve the social welfare,
we have proposed a deep reinforcement learning based auction (DRLA)
mechanism which uses the graph neural network to effectively extract
useful features from worker's reported types and automatically determines
the service allocation and payment. Both the proposed RMA mechanism
and the DRLA mechanism possess the economic properties of truthfulness
and individual rationality. 

\bibliographystyle{IEEEtran}
\bibliography{PJ7_FL_AUCTION}
\end{document}